\documentclass[a4paper,12pt]{article}
\pdfoutput=1
\usepackage{graphicx, rotating,amssymb,amsmath}
\usepackage{epstopdf,pdfsync}

\usepackage{cite}
 \usepackage{hyperref}

\hypersetup{colorlinks,bookmarksopen,bookmarksnumbered,citecolor=verdeCP3,
linkcolor=bluCP3,pdfstartview=FitH,urlcolor=rossoCP3}

\def\mhref#1{\href{mailto:#1}{#1}}		

\usepackage{slashed}
\usepackage{multicol}
\usepackage{color}
\definecolor{rossoCP3}{cmyk}{0,.88,.77,.40}
\definecolor{verdeCP3}{rgb}{0.09765625, 0.57421875, 0.1015625}
\definecolor{bluCP3}{rgb}{0, 0.23, 0.67}

\def\circa#1{\,\raise.3ex\hbox{$#1$\kern-.75em\lower1ex\hbox{$\sim$}}\,}
\makeatletter

\long\def\symbolfootnote[#1]#2{\begingroup\def\thefootnote{\fnsymbol{footnote}}
\footnote[#1]{#2}\endgroup}

\begin{document}

\begin{titlepage}

\begin{flushright}
\scriptsize
CP$^3$-Origins-2015-016 DNRF90 
\end{flushright}
\color{black}

\vspace{0.3cm}

\begin{center}
{\Large {\color{rossoCP3}\bf First Order Electroweak Phase Transition \\ ~\\ from  \\ ~\\ (Non)Conformal Extensions of the Standard Model}}
\end{center}

\par \vskip .4in \noindent

\begin{center}
{\sc Francesco Sannino$\, {\color{bluCP3}^{1}}$\,\symbolfootnote[1]{\mhref{sannino@cp3-origins.net}},
Jussi Virkaj\"arvi$\, {\color{bluCP3}^{1,2}}$\,\symbolfootnote[5]{\mhref{virkajarvi@cp3-origins.net}}}
\end{center}

\begin{center}
\par \vskip .2in \noindent
 {\small ${\color{bluCP3}^1} \, $\href{http://cp3-origins.dk}{CP$\,^3$-Origins}, Univ. of Southern Denmark, 
Campusvej 55, DK-5230 Odense M, Denmark}
\\~\\
{ \small ${\color{bluCP3}^2} \, $\href{https://www.jyu.fi/fysiikka/en}{Dept. of Physics}, Univ. of Jyv\"askyl\"a,
P.O.Box 35, FI-40014, University of Jyvaskyla, Finland}
\\

\par \vskip .3in \noindent

\end{center}

\begin{center}
{\large Abstract}
\end{center}

\begin{quote}
We analyse and compare the finite-temperature electroweak phase transition properties of classically (non)conformal extensions of the Standard Model. In the classically conformal scenarios the breaking of the electroweak symmetry is generated radiatively. The models feature new scalars coupled conformally to the Higgs sector as well as new fermions. We uncover the parameter space leading to a first-order phase transition with(out) the Veltman conditions. 
We also discuss dark (matter) aspects of some of the models and compare with existing literature when appropriate. We observe that to accommodate both, a first order electroweak phase transition, and a phenomenologically viable dark matter candidate requires to go beyond the simplest extensions of the Standard Model. Furthermore these extensions must all feature new degrees of freedom that are naturally lighter than a TeV and therefore the associated models are testable at the upcoming Large Hadron Collider run two experiments. 
\end{quote}

\par \vskip .1in \noindent


\end{titlepage}

\newpage

\tableofcontents

\newpage


\section{Introduction}
\label{intro} 
The Standard Model (SM) is one of the most successful theories of nature. It is nevertheless incomplete and suffers of a numbers of theoretical shortcomings. For example it is unable to account for the dark matter (DM) problem  and cannot explain baryogengesis. 

It is therefore relevant to investigate minimal extensions that can also address some of its flaws. In particular we would like to investigated extensions that can lead to a strong first order electroweak phase transition (EWPT). 

We know, in fact, that any model attempting to address the baryon asymmetry in the Universe should fulfill the three Sakharov's conditions \cite{Sakharov:1967dj}: The presence of baryon number violating operators; C and CP violations and out of equilibrium conditions. The reasons why the SM cannot accommodate a baryon asymmetry is that the associated EWPT is a smooth cross over, for the physical Higgs boson mass value \cite{Kajantie:1996mn}. Furthermore the SM is not a source of sufficiently large CP violation that is needed to generate the observed particle-antiparticle asymmetry. 
 
If we insist on having an EW scale induced baryogenesis we must extend the SM in such a way that the Sakharov conditions are met. Here we investigate in particular the last condition that requires the presence of a sufficiently strong first order EWPT.  

Because of the vast landscape of possible extensions  of the SM we  concentrate on minimal ones. We consider, therefore,  extensions featuring new scalars and fermions and explore in detail the associated classically conformal limit where the electroweak symmetry breaks radiatively at zero temperature. 
{For each model we present first the non-conformal classical limit and then move to investigate in detail the classically conformal limit with(out) imposing the Veltman conditions (VC).}

We unveil the region in coupling space allowing for a first order EWPT and, when relevant, we also discuss the dark (matter) aspects. Not surprisingly we observe that to accommodate both, a first order EWPT, and a phenomenologically viable DM candidate requires to go beyond the simplest extensions of the SM.  Interestingly, however, the extensions that can lead to a first order EWPT must feature new degrees of freedom that are naturally lighter than a TeV. This fact renders the models primary candidates to be tested at the second run of the Large Hadron Collider (LHC) experiment. Moreover requiring the models to be classically conformal and/or to satisfy the one-loop VCs strongly reduces the available parameter space of the couplings substantially increasing their productivity.  

To analyse the finite-temperature properties we use the perturbative approach which we now briefly summarise. We study the EWPT using the ring improved 1-loop finite temperature effective potential given in Eq.~\ref{potential} of App.~\ref{Effpot}\footnote{Except, in Sec.~\ref{case2} where we use the tree level potential augmented only by the leading finite temperature corrections  \cite{Espinosa:2011ax}.}. We then solve for the quantity $h_c  /T_c$. As indication of the possible occurrence of a strong first order phase transition we require $h_c  /T_c \gtrsim 1$.  Here $h_c \equiv \langle h(T_c) \rangle  \neq 0$ is the vacuum expectation value (VEV) of the Higgs field, in the EW symmetry broken minimum of the potential, at the critical temperature $T_c$. The quantities, $ h_c $ and $T_c$, are determined from the potential when two degenerate minima appear in the potential separated by an energy-density barrier.

The different degrees of freedom (DoF) that contribute to the effective potential, Eq.~\ref{potential}, are given in App.~\ref{SMmass}, \ref{AppHS} and \ref{Appsxy} for each model separately. In all our calculations we use the Landau gauge, and for each model from the SM side we consider corrections from the gauge bosons, Z and W, and the top quark. 
Following common practice (e.g. \cite{Espinosa:2008kw,Cline:2009sn}), we
neglect the the Goldstone boson contributions to the effective potential\footnote{To improve the analysis one could use a resummation technic for these states following \cite{Elias-Miro:2014pca,Martin:2014bca}.}. Similarly, in the conformal case we neglect the Higgs boson contribution.

 In Sec.~\ref{SM} we summarise the EWPT status for the SM. To set the stage, and to acquire insight  we also consider the EWPT for the SM with imposed VCs and/or its classically conformal limit even though these models are ruled out by experiments. 
 
 We then move to Sec.~\ref{Sec3SMS} and Sec.~\ref{SMCS} where we systematically extend the SM and investigate the order of the associated EWPT.    When relevant we also discuss the DM properties of the models. We finally offer our conclusions in  Sec.~\ref{conc}. 
 We provide the bulk of the  computations, including the expressions for the  effective potential at nonzero temperature, for each model   in several appendices App.~\ref{Effpot}-\ref{AppAnn}.


\section{(Non)conformal Standard Model}
\label{SM}
We start our investigation by briefly summarising the EWPT status in the SM. For illustration and to set the stage for future examples we also investigate the model with imposing the Veltman conditions \cite{Veltman:1980mj} as well as  the limit in which the symmetry breaks radiatively via a Coleman-Weinberg phenomenon \cite{Coleman:1973jx}.   
 
\subsection{Standard Model} 
 
The EWPT within the minimal SM has been studied extensively in the past using perturbative methods \cite{Arnold:1992rz,Fodor:1994bs,Farakos:1994kx} and non-perturbative lattice simulations \cite{Kajantie:1996mn,Rummukainen:1998as,Laine:1998jb,Csikor:1998eu,Kajantie:1995kf,Kajantie:1996qd,Philipsen:1996af,Gurtler:1997hr,Karsch:1996yh}. 
Based on these studies, it is well known, that the EWPT in the SM is not first order but a smooth cross over, for the physical Higgs boson mass value. More specifically one finds, via lattice simulations, that the Higgs mass should be less than about  70 GeV for the phase transition to be first order.

 We shall be using the perturbative analysis to provide a first qualitative understanding of the phase structure of several extensions of the SM. 
 This analysis becomes robust in the limit when the thermal-loop corrections are reliable.
 This occurs for certain region of the parameter space of different theories. In the SM this occurs for light Higgs masses, i.e.~less than 70 GeV. 
The break down of the perturbative analysis, for heavier Higgs boson masses in the SM, is due to the so-called Linde's problem \cite{Linde:1980ts} which is  caused by the gauge bosons that are light near the EW symmetric phase. However in models with extra scalars coupled to the Higgs via a portal coupling they can provide an important contribution to the transition, de facto, improving on the reliability range of the perturbative analysis. 
We exemplify this issue below.

To elucidate the potential mechanism that can yield a first order nature of the electroweak phase transition we consider the simple case when the SM  effective potential can be approximated to be
\begin{eqnarray}
\label{EffSM}
V^{\rm eff}_{ \rm SM}(h,T) \approx \frac{1}{2} m^2_{\rm eff}(T) h^2 - E T h^3 +  \frac{1}{4}  \lambda_{\rm eff} h^4  \,,
\end{eqnarray}
where $m^2_{\rm eff}(T)= -\mu^2 + \frac{1}{16}(3g^2+g'^2+4y^2_t+8\lambda)T^2$ is the thermally corrected Higgs boson mass term giving the leading finite temperature term of the potential $\sim T^2 h^2$. 
The bosonic degrees of freedom provide the second term to the effective potential, $- E T h^3$, where $E$ is a coupling constant dependent function. More precisely, if taken into account only the most relevant gauge boson contributions, $E= \frac{1}{12\pi}(6(g/2)^3+3(\sqrt{g^2+g'^2}/2)^3)$. This $- E T h^3$ term plays an important role for a possible first order nature of the EWPT\footnote{This
same bosonic term provides the 1st order EWPT also in the conformal Higgs portal models studied in this work. In these models it is usually the Higgs portal coupling $\lambda_{hs}$ which dominates the $E$ factor and thus also limits the so called Linde's problem.}.
This follows from the fact that it is negative and it has a cubic dependence on the background Higgs field. Indeed, the interplay of this term with the positive leading term $\sim T^2 h^2$, and the positive quartic term $\frac{1}{4} \lambda_{\rm eff} h^4$, where $ \lambda_{\rm eff}(T) \approx \lambda$  allows for the formation of two degenerate minima in the effective potential. 
{The barrier separating the two vacua hints to a first order nature of the transition. And indeed the transition is first order, as long as the Higgs mass is $\lesssim 70$ GeV \cite{Kajantie:1996mn}. However, the transition is not strong, i.e.~$h_c  /T_c \lesssim 1$, for Higgs mass values larger than $\sim 10$ GeV.}

Note that, even though the approximative effective potential, Eq.~\ref{EffSM}, qualitatively describes the EWPT at low Higgs masses, it is not a proper analysis.
Indeed, if we use Eq.~\ref{EffSM} and the dominant W-boson contributions in the $E$ factor, we get that $h_c/T_c \approx 2E/ \lambda \approx g\, m^2_W/ ( \pi \, m^2 _H )$.
However, this approximation is not reliable particularly for heavy Higgs boson masses. This is due to the previously mentioned Linde's problem, that occurs when the gauge boson mass, $m \sim g \,h$, is of the order of the magnetic mass, $m \sim g^2 T$. In this case the high temperature expansion parameter, of the form $ \sim g^2 T/ m$, becomes of the order of unity, and the perturbative approximation breaks down. 

We can circumvent the problem by requiring $g h \gtrsim g^2 T$ near the transition, which yields $h/T \gtrsim g$. 
On the other hand, we had that $h_c/T_c \sim g \, m^2_W/m^2_H$. Thus, the perturbative results holds as long as  $m_W/m_H \gtrsim 1$, where as for $h_c  /T_c \lesssim 1$ the perturbative results are less reliable. 
Finally, notice also that, if the daisy diagram resummation\footnote{The daisy diagram resummation procedure is introduced to cure the IR divergences related to the bosonic zero modes \cite{Carrington:1991hz,Arnold:1992rz}.}, affecting the $- E T h^3$ term,  is taken into account, it makes the transition slightly weaker. 
On the other hand if higher order corrections are taken into account they tend to make the transition stronger \cite{Arnold:1992rz}.     

Because we will be using the perturbative analysis here, other non-perturbative methods should also be employed to firmly establish the nature of the phase transition.  We have checked our 1-loop SM result for $h_c /T_c$ against the one of Ref.~\cite{Arnold:1992rz} in Figs. 25 and 27, and found a very good agreement. 


\subsubsection{Veltman conditions}

The scalar boson(s) suffer from the naturality problem expressed, for the SM, in terms of the presence of the quadratic divergences for the Higgs boson mass operator. To alleviate this problem, one can impose the VCs \cite{Veltman:1980mj} for the scalar mass term.  In practice this requires to set a relation between the model couplings at each order in perturbation theory, so that the quadratically divergent piece cancels.

Using a cutoff regularisation at the one loop level in the SM the quadratically divergent piece reads
\begin{eqnarray}
\label{deltamH2}
\delta_{\Lambda^2} m^2_H = \left[ 6 \lambda+\frac{9}{4}g^2+\frac{3}{4}g'^2-6y^2_t \right] \frac{ \Lambda^2}{16 \pi^2}  \,.
\end{eqnarray}
To this order the VC  requires $\delta_{\Lambda^2}  m^2_H = 0$\footnote{
Here $\delta_{\Lambda^2} $ indicates that we are only taking into account mass corrections which are proportional to ${\Lambda^2}$, leaving out the $\log({\Lambda^2})$-terms. Further, it should be understood, here and here after, that all the couplings are defined at the same renormalization point (scale) $Q$, so that e.g.~weak coupling $g \equiv g(Q)$ has the measured value at the fixed scale, for example, at the Z boson mass scale $Q=m_Z$.} yielding 
\begin{eqnarray}
\label{VeltSM}
6 \lambda+\frac{9}{4}g^2+\frac{3}{4}g'^2-6y^2_t =0  \,.
\end{eqnarray}
It is well known that this constraint is not satisfied phenomenologically for the SM. Nevertheless for illustration we show its consequences for the EWPT. We keep the gauge boson masses, i.e.~$g$ and $g'$ coupling values, as measured and use the VC to either predict the wrong Higgs boson mass or the wrong top quark mass i.e.~the associated coupling value for $\lambda$ or for the  top Yukawa $y_t$.

If we keep the correct top mass value, i.e.~$m_t \approx 173$ GeV, the VC, Eq.~\ref{VeltSM} yields  
$m_H \approx 314$ GeV. Thus, in this case, since the Higgs boson is very heavy, perturbation theory is not reliable, and as previously discussed, lattice results show that the EWPT is a cross over.

If, on the other hand, we keep the Higgs boson mass as measured $m_H \approx 126$ GeV the top mass gets fixed to  $m_t \approx 96$ GeV. Using perturbation theory, we deduce $h_c /T_c \approx 0.1$, which does not suggest the onset of a strong first order EWPT.
{For such a small ratio, and because the Higgs mass is large, i.e.~larger than 70 GeV, we resort to lattice field theory that, as in the case of the SM, predicts the EWPT  to be a cross over. }
 
To summarize, the VCs do not alter the fate of the EWPT in the SM, i.e. it remains a cross over.


\subsection{Conformal Standard Model} 
 
Now we consider EWPT in the classical conformal limit of the SM. This allows us to investigate how different assumptions on the couplings and structure of the model affect the EWPT.
 
 In this case the EW symmetry breaking is triggered by the CW mechanism \cite{Coleman:1973jx}. Here the tree level potential $V(h)=\frac{\lambda}{4} h^4$ needs to be flat i.e.~it needs to fulfil the condition $\lambda \approx 0$ at the electroweak scale.   The CW prediction for the one-loop generated Higgs mass is 
\begin{eqnarray}
\label{Higgsmass}
m^2_H = \frac{3}{8 \pi^2} \left[ \frac{1}{16}(3 g^4+2g^2 g'^2+g'^4) +4 \lambda^2-y^4_t \right] v^2_{EW} \,,
\end{eqnarray}
where $v_{EW} = 246$ GeV is the VEV of the Higgs field. For more details see e.g.~App.~\ref{flat}.

Using the flatness condition, $\lambda \approx 0$, and assuming the correct top quark mass gives $m^2_H < 0$, which leads to a further instability, that can be avoided by reducing the (Yukawa $y_t$) quark mass to smaller  unphysical values.  By using Eq.~\ref{Higgsmass} with $\lambda \approx 0$ and floating $y_t$, we get that $0 \lesssim m_H \lesssim 9.8$ GeV with top mass varying in the range $78.6 \gtrsim m_t  \gtrsim 0$ GeV respectively. Actually the lower (upper) bound for the top quark (Higgs boson) mass should not be taken strictly because  we have neglected all the other fermions.   Thus for example by fixing $m_t \approx 10$ GeV we have $m_H \approx 9.8$ GeV and $h_c /T_c \approx 8$, which indicates  the occurrence of a very strong first order phase transition. 
Increasing the top quark mass value, and thus also lowering the Higgs boson mass value, makes the transition even stronger.

One should notice however, that, for large values of $h_c / T_c $, e.g.~$h_c / T_c \gtrsim 5$, the energy density
barrier between the EW symmetric vacuum and one of the broken minima may be so large, that the symmetric phase becomes meta-stable and the model becomes not viable. 
Since, however,  the potential barrier in the classically conformal scenario is only radiatively induced at nonzero temperature, (and absent at one loop at zero temperature) the metastable vacuum could decay. 
In the other models, studied in this work, we restrict our attention to the cases $h_c / T_c \geq 1 $ but with the ratio which is not too large $h_c / T_c  \leq 4 $. 
This is typically sufficient to avoid the meta-stability problem.  Therefore we do not perform         
 in depth meta-stability analysis \cite{Coleman:1977py,Callan:1977pt,Linde:1980tt,Linde:1981zj}. 


\subsubsection{Veltman conditions}
 Classical conformality does not address the naturality problem as discussed in \cite{Antipin:2013exa}. A way to alleviate this problem also for the classical conformal scenario is to further impose the VC  \cite{Antipin:2013exa}. 

By imposing the VC on the conformal SM limit at the one-loop level, Eq.~\ref{VeltSM} predicts that the top (Yukawa) quark mass should be  $m_t \approx 73$ GeV and $m_H \approx 5$ GeV  as shown in \cite{Antipin:2013exa}.
Because the top is relatively heavy, and the EW symmetry breaking occurs radiatively, the EWPT is very strongly first order as indicated by the $h_c /T_c \approx 16$  value.
We remind the reader of a potential meta-stability issue \cite{Coleman:1977py,Callan:1977pt,Linde:1980tt,Linde:1981zj}.

To summarize, the SM does not support a 1st order EWPT with or without VC.  The situation changes dramatically in the case of the classically conformal SM where a strong first order phase transition occurs with and without VC albeit the model is not phenomenologically viable. 

Thus, we would like to  investigate and compare (classically conformal) extensions of the SM with(out) VC that can support a 1st order EWPT.

A reasonable strategy is to search for models that enhance the negative cubic term $- E T h^3$. As we shall see this mechanism is particularly efficient for the conformal models since they have a flat direction for the effective potential at the tree-level. The enhancement is then achieved by adding new bosonic degrees of freedom to the model. This guides us towards theories containing new scalar degrees of freedom to be added to the SM. 


\section{(Non)conformal Extension with A Real Singlet Scalar} 
\label{Sec3SMS}

We begin the investigation of extensions of the SM that can support a first order EWPT with Higgs portal models featuring a real singlet scalar. This is a heavily studied model and it is reported here mostly for completeness. We will refer to the existing literature when needed while concentrating particularly on the classical conformal limit with(out) VC.  

\subsection{SM + Real Singlet Scalar} 
\label{SMrS}

 The most general renormalizable tree level scalar potential for this model reads:
\begin{eqnarray}
\label{ScalarsSMS}
V_{0}& =& -\mu^2 H^{\dagger}H  + \lambda_h (H^{\dagger}H)^2 +  \frac{m_S^2}{2}S^2 + \lambda_{hs}H^{\dagger}HS^2+  \frac{\lambda_{s} }{4}  S^4 + \nonumber \\
          &+ &  \frac{\mu_{3}}{3}S^3+\mu^3_{1}S+\mu_{m}SH^{\dagger}H \ , 
 \end{eqnarray}
where $S$ is the new real singlet scalar. The  last three terms  are absent if one requires the potential to be $Z_2$ symmetric. The finite-temperature phase structure can be studied by replacing the fields in the potential by their background values $h$ and $s$. The associated tree level potential is obtained by substituting $H \rightarrow h/ \sqrt{2}$ and $S \rightarrow s$ in Eq.~\ref{ScalarsSMS}. 
Depending on the values of the model parameters the phase structure and the vacuum structure can be very rich. We shall not make a full EWPT analysis for this general model here, as it can be found from the literature \cite{Choi:1993cv,Ham:2004cf,Ahriche:2007jp,Profumo:2007wc,Ashoorioon:2009nf,Espinosa:2011ax} (see also \cite{Ahriche:2012ei}). A thorough investigation of the potential structure, yielding a strong first order phase transition stemming already from the tree level potential, once the leading finite temperature terms are taken into account, was recently performed in \cite{Espinosa:2011ax}.  We will consider this kind of scenario later in Sec.~\ref{case2}. 

The EW symmetry breaking and first order phase transition can also be obtained via loop corrections. This is argued  in Sec.~\ref{cSMrS} where we will discuss the EWPT for the classical conformal case with(out) VC.

If the $Z_2$ symmetry is imposed  the model can also feature a DM candidate because the singlet becomes stable. The DM phenomenology and EWPT of this type scenario has been studied recently e.g.~in \cite{Cline:2012hg,Cline:2013gha}.  The authors concluded that in order to have a sufficiently strong first order EWPT capable to produce baryogenesis, the singlet can only be a subdominant DM component.


\subsection{Conformal SM +  Real Singlet Scalar} 
\label{cSMrS}

In the classically conformal case the tree level scalar potential reads:
\begin{eqnarray}
\label{ScalarsSMSZ2}
V_{0} =  \lambda_h (H^{\dagger}H)^2 + \lambda_{hs}H^{\dagger}HS^2+  \frac{\lambda_{s} }{4} S^4\,,
\end{eqnarray}
which automatically respects the $Z_2$-symmetry, as no terms with dimensional couplings, including the terms with odd number of $S$ fields, are allowed by the classical conformality requirement \footnote{Of course classical conformality per se does not always imply the presence of a $Z_2$-symmetry. This is apparent from the case studied in Sec.~\ref{PNC}, where a new singlet Weyl fermion, with a classically conformal tree level interaction term $y_{\chi} S \chi \chi +h.c.$, is added to the model given in Eq.~\ref{ScalarsSMSZ2}.  Therefore unless a $Z_2$-symmetry is strictly enforced, the scalar $S$ does not need to be stable, even though the tree level scalar potential possesses an accidental $Z_2$-symmetry. Furthermore if $S$ acquires a VEV one should expect the presence of new sectors breaking explicitly the $Z_2$-symmetry to avoid the domain wall problem \cite{Espinosa:2011eu,Zeldovich:1974uw}.}.
Stability considerations constrain the model parameters to be 
\begin{eqnarray}
\label{SMScoup}
\lambda_h \geq 0 \,\, \rm{and}  \,\,  \lambda_s \geq 0 \,, \, \,\rm{and} \, \,\rm{if}\, \, \lambda_{hs} < 0 \,,  \,\rm{then} \, \,  \lambda^2_{hs}  \leq \lambda_h \lambda_s \,.
\end{eqnarray}
In this case the EW symmetry breaking follows from the CW mechanism. Because the CW mechanism, to be operative, requires the presence of flat directions at the electroweak scale we can consider different scenarios. Here we shall consider two cases. The case (1) in which the coupling $\lambda_{hs}$ is positive and the flat direction is in the $h$-direction. In this case  $S$ does not acquire a non-zero VEV, i.e.~$\langle s \rangle = 0$ at zero temperature (and at finite temperature). In case (2), $\lambda_{hs}$ is negative, and the flat direction is in some non-trivial direction in the (h,s)-plane. Here the field $S$ acquires a non-zero VEV, i.e.~$\langle s \rangle \neq 0$ at zero temperature (and at finite temperature).


\subsubsection{Case (1): $\langle s \rangle =$ 0} 
\label{SMrSc1}

For $\lambda_{hs}$  positive and with the scalar quartic couplings satisfying the relation $\lambda_{s} >\lambda_{h} = 0$, the $S$ field does not acquire a VEV and the flat direction can be taken to be in the Higgs direction.
 
The phase transition for this type of model has been investigated previously in \cite{Espinosa:2007qk,Espinosa:2008kw}.  There, however,  the singlet quartic self coupling effect was not included but on the other hand the model included several real singlet scalars with identical portal couplings $\lambda_{hs}$.  We were able to reproduce their results  and find 
 that in the case of 12 real singlet scalars, the EWPT can be strongly first order with $h_c /T_c \approx 2.7$, provided the portal coupling is $\lambda_{hs} \approx 1.4$\footnote{The notation for the portal coupling in \cite{Espinosa:2007qk,Espinosa:2008kw} is different from ours, the two are related via  $\lambda_{hs} = \xi^2$.}.

We used the full one-loop effective potential, Eq.~\ref{potential}, with mass parameters, as defined in App.~\ref{AppHS}.
The flat direction condition is $\lambda_h(Q) \approx 0$, with $Q \approx v_{EW}$, and the CW analysis is similar to the one for the SM but able to provide the correct Higgs boson mass $\approx 126$ GeV. 
In the case of only one singlet scalar, and neglecting $\lambda_{s}$, we find that $h_c /T_c \approx 1.76$ for  
$\lambda_{hs} \approx 4.84$ which also allow for the correct Higgs boson mass. The singlet tree level mass, follows from the relation $m^2_s = \lambda_{hs}  v^2_{EW}$ and yields $m_s \approx 541$ GeV. 

\begin{figure}
\begin{center}
\includegraphics[width=0.49\textwidth]{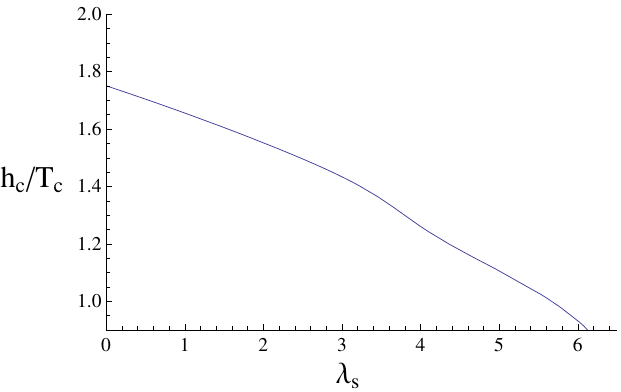} 
\caption{Plot of $h_c /T_c$ as a function of $\lambda_{s}$ for the real singlet Higgs portal model of case (1). Here the Higgs boson has mass $126$ GeV and the singlet $S$ has mass 541 GeV.
}
\label{fig:SMS_c1}
\end{center}
\end{figure}

We show now how the inclusion of $\lambda_s$ affects the above results. The one-loop Higgs mass and the tree level singlet masses are unaffected by the addition of the $S$ quartic interaction term. $\lambda_s$ affects the EWPT because it appears in the singlet scalar thermal mass. Increasing  $\lambda_s$  decreases the ratio $h_c /T_c $. 
{This follows from the fact that the temperature dependent piece, in the daisy improved boson-induced cubic term in the effective potential increases with $\lambda_s$. This damps the field dependent piece responsible for the 1st order nature of the transition.
In turn, one observes a decrease of the ratio $h_c /T_c $, as it can be seen from Fig.~\ref{fig:SMS_c1}.}
 
{Finally, using the DM results from \cite{Cline:2013gha}, we conclude that in the conformal case the singlet can provide at most a subdominant DM component with density $\lesssim 1\%$ of the observed DM density.}

For the Higgs portal models we allow relatively large couplings, i.e.~up to $\le 4 \pi$. Of course, if some of the parameters approach $ 4 \pi$ it is clear that the one-loop approximation is not reliable. In this case one should take into account higher loop effects, or better use directly non-perturbative methods. However, recently in \cite{Tamarit:2014dua}, for complex singlet Higgs portal models it has been shown, by an explicit 2-loop computation, that identifying a 1st order EWPT via the 1-loop effective potential provides reliable results up to quite large values of the couplings of the order of $\lambda_{hs} \sim 8$.


\subsubsection{Veltman conditions}

Now we impose the VC for this model. In this case we have two conditions, one for each scalar field.
The singlet scalar VC is defined similarly as the Higgs one, i.e. $\delta_{\Lambda^2}  m^2_S = 0$. The conditions read for $S$ and $H$ respectively \cite{Antipin:2013exa}
\begin{eqnarray}
\label{eqVCS}
3 \lambda_s+4 \lambda_{hs} = 0  \ ,
\end{eqnarray}
 \begin{eqnarray}
\label{eqVCH}
6 \lambda_h+\frac{9}{4}g^2+\frac{3}{4}g'^2-6y^2_t + \lambda_{hs}=0  \,.
\end{eqnarray}
We observe immediately from Eq.~\ref{eqVCS} that $\lambda_{hs}$ should be negative since $\lambda_s \geq 0$ to ensure stability of the potential. However, the EWPT study and CW analysis, in Sec.~\ref{SMrSc1} above, indicated, that $\lambda_{hs}$ should be positive,  to be able to support a strong first order EWPT and the correct Higgs boson mass value. Thus the VCs are naturally applicable to case (2) rather than (1) for which $\lambda_{hs} < 0$ from the start.


\subsubsection{Case (2): $\langle s \rangle \neq$ 0} 
\label{SMS_c2}

{ The flat direction is not in the Higgs direction but it is chosen to be along some non-trivial direction in the (h,s)-plane.}
 For the CW analysis we follow \cite{Antipin:2013exa} and review the flat direction conditions in App.~\ref{flat}.

It is convenient to write the potential, Eq.~\ref{ScalarsSMSZ2}, using polar coordinates, i.e.~$H=(0, r \cos \omega)^T / \sqrt{2}$ and $S=r \sin \omega$ which now reads: 
\begin{eqnarray}
V_{0}& =& \frac{r^4}{4}(\lambda_h \cos^4 \omega + \lambda_s \sin^4 \omega+ 2\lambda_{hs} \sin^2 \omega \cos^2 \omega) \ .
\end{eqnarray}
The minimum of the potential, along some ray $r$, is found by studying the first and second derivatives of the potential, leading to  the conditions
\begin{eqnarray}
0 \leq \lambda_h  < \min \{\lambda_s, \lambda_{hs} \} \, : && \langle \omega \rangle = 0 \ , \label{eqflatSM} \\
0 \leq  \lambda_s < \min \{\lambda_h, \lambda_{hs} \} \,: &&\langle \omega \rangle = \frac{\pi}{2} \ , \label{eqflatS}\\
-\sqrt{\lambda_h \lambda_{s}}  \leq \lambda_{hs} < \min \{\lambda_h, \lambda_{s} \} \, : &&\tan^2 \langle \omega \rangle = \frac{\lambda_h- \lambda_{hs}}{\lambda_s- \lambda_{hs}} \ . \label{eqflatnontri}
\end{eqnarray}
We chose to use polar coordinates, because
is it  very useful  for the CW and EWPT analysis, as we will now discuss.
Indeed, once the model parameters $\lambda_s, \lambda_h$ and $\lambda_{hs}$ are such, that the flat direction condition is fulfilled, then the fixed angle $\langle \omega \rangle $ defines simultaneously the minimum direction and the flat direction of the tree level potential, while $\langle r \rangle $ the location of the ground state along the flat direction. Thus once this direction is fixed to $\langle \omega \rangle $, the CW and EWPT analyses can be performed directly along the ray $r$, see App.~\ref{flat} for a detailed discussion of the method. 
 
Now, the first case, $\langle \omega \rangle = 0 $, given in Eq.~\ref{eqflatSM} above, corresponds to  the case (1) studied in Sec.~\ref{SMrSc1} above.
On the other hand for $\langle \omega \rangle = \frac{\pi}{2} $, given in Eq.~\ref{eqflatS}, the EW symmetry does not break and we are left with the third possibility, i.e. Eq.~\ref{eqflatnontri}. The flat direction condition in this case reads 
\begin{eqnarray}
\label{flatc0}
\lambda_h \lambda_s - \lambda^2_{hs}=0 \ .
\end{eqnarray}
The CW mechanism along the flat direction,  Eqs.~\ref{flatc0} and \ref{eqflatnontri}, leads to spontaneous symmetry breaking with $\langle r \rangle \cos \langle \omega \rangle = v_{EW} $. The flatness and VCs are simultaneously satisfied at $Q \sim v_{EW}$. By expanding our fields around the ground state via $r \cos \langle \omega \rangle = v_{EW} + h_0$ and  $r \sin \langle \omega \rangle = v_{EW} \tan \langle \omega \rangle + s_0$,
we find the mass eigenstate fields 
\begin{eqnarray}
\phi= h_0 \cos \langle \omega \rangle + s_0\sin \langle \omega \rangle \ , \quad \Phi= s_0 \cos \langle \omega \rangle - h_0 \sin \langle \omega \rangle \,,
\end{eqnarray}
with tree level masses $m^2_{0,\phi} =0$ and $m^2_{0,\Phi} =2(\lambda_h-\lambda_{hs}) v^2_{EW}$. Here $\phi$  is the physical Higgs field and $\Phi$ the heavier state.
The CW induced one-loop mass for the Higgs field is 
\begin{eqnarray}
\label{HiggsmassS1}
m^2_{\phi} = \frac{\cos^2  \langle \omega \rangle}{8 \pi^2} \left[ \frac{1}{16}(6 g^4+3(g^2 +g'^2)^2) +4 (\lambda_h-\lambda_{hs})^2- \frac{12}{4}y^4_t \right] v^2_{EW} \,.
\end{eqnarray}
We can now constrain two out of the three model parameters $\lambda_h, \lambda_s$ and $\lambda_{hs}$, using the flat direction condition, Eq.~\ref{flatc0}, and the Higgs boson mass relation, Eq.~\ref{HiggsmassS1}.  We choose to first fix $\lambda_{hs}$ using the flat direction condition, Eq.~\ref{flatc0}.
By taking into account the stability conditions $\lambda_{h,s} \geq 0$ and the minimum condition, Eq.~\ref{eqflatnontri}, this gives $\lambda_{hs} = -\sqrt{\lambda_{h} \lambda_{s}}$. Then, using the previous relation, 
as well as the measured $y_t, g$ and $g'$ coupling values, in the Eq.~\ref{HiggsmassS1}, we can search for real valued solutions for $\lambda_{h}$, as a function of $\lambda_{s}$, providing the physical Higgs boson mass value.
Given that $\lambda_{hs}$ and $\lambda_{h}$ are now functions of $\lambda_{s}$, 
we then performed a scan over the last remaining free parameter $\lambda_{s}$, in our EWPT study within the range 
\begin{eqnarray}
 0.01 \leq \lambda_s < 4 \pi \,, 
 \end{eqnarray}
with a flat prior distribution  with 1000 points.
In the phase transition analysis we used the 1-loop effective potential, Eq.~\ref{potential}, with the relevant mass terms given in App.~\ref{AppHS} and with the CW conditions discussed above and given also in App.~\ref{flat}.

\begin{figure}[t]
\begin{center}
\includegraphics[width=0.30\textwidth]{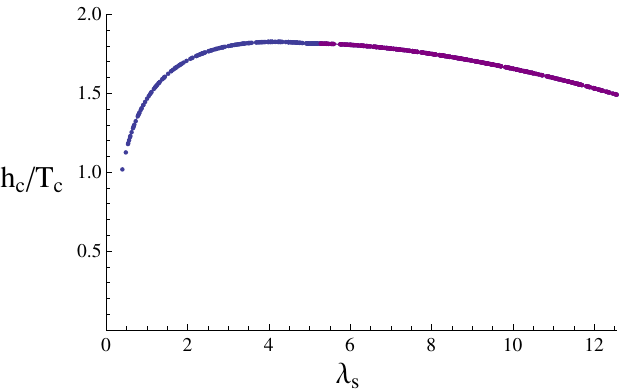} 
\includegraphics[width=0.30\textwidth]{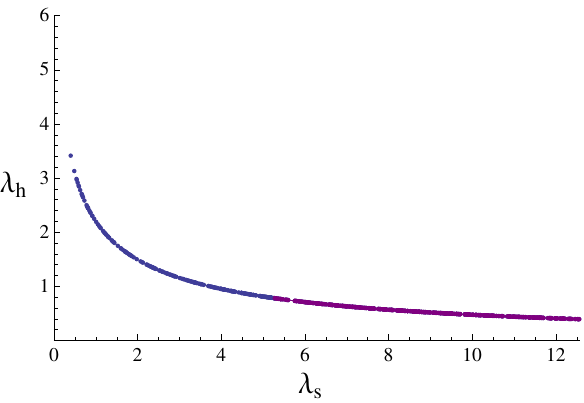} 
\includegraphics[width=0.30\textwidth]{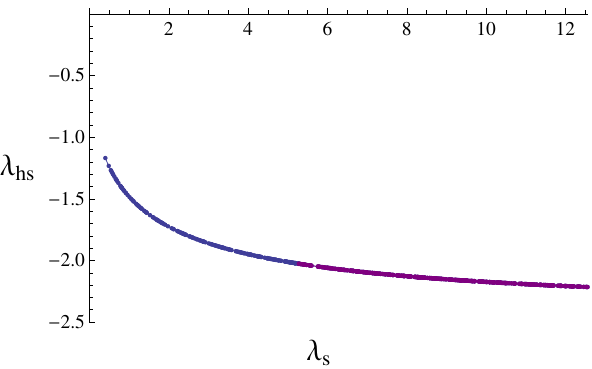} 
\includegraphics[width=0.30\textwidth]{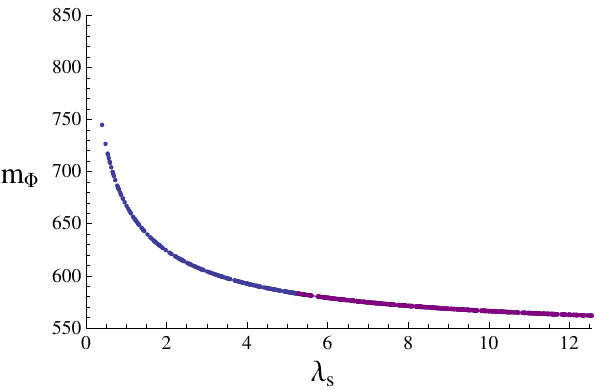} 
\includegraphics[width=0.30\textwidth]{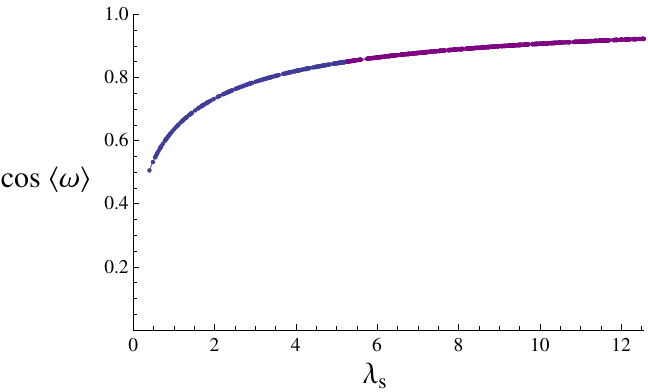} 
\caption{Real singlet Higgs portal model parameters as a function of $\lambda_s$ for case (2) that yield a first order EW phase transition. The purple (blue) points with(out) the constraint $\cos \langle \omega \rangle \geq 0.85 $. For all the model points the Higgs boson mass assumes the correct value. The model parameters have not been constrained using the VCs.}
\label{fig:SMS_c2_ls}
\end{center}
\end{figure}

\begin{figure}[t]
\begin{center}
\includegraphics[width=0.30\textwidth]{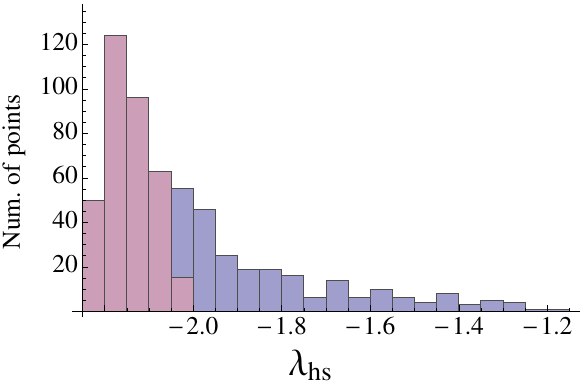} 
\includegraphics[width=0.30\textwidth]{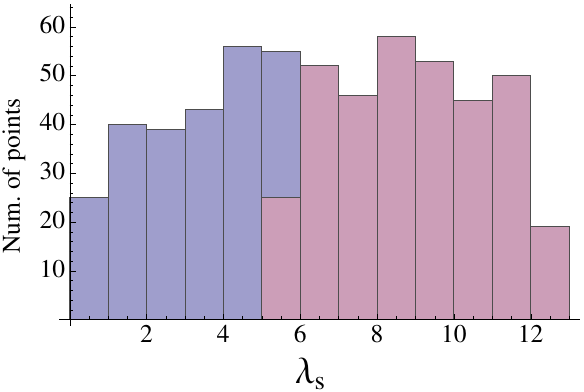} 
\includegraphics[width=0.30\textwidth]{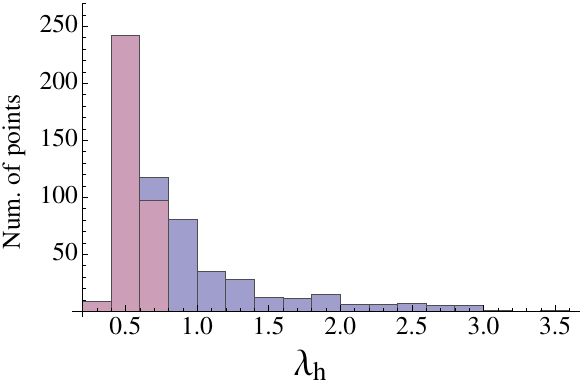} 
\includegraphics[width=0.30\textwidth]{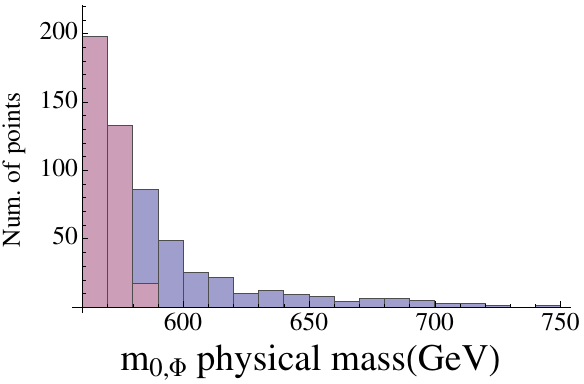} 
\includegraphics[width=0.30\textwidth]{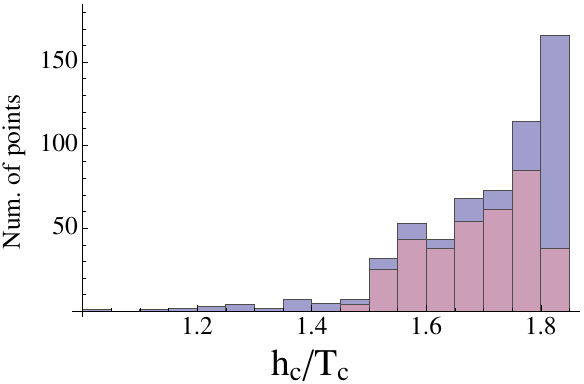} 
\includegraphics[width=0.30\textwidth]{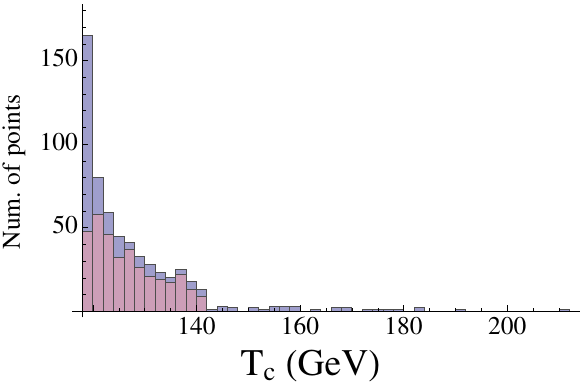} 
\includegraphics[width=0.30\textwidth]{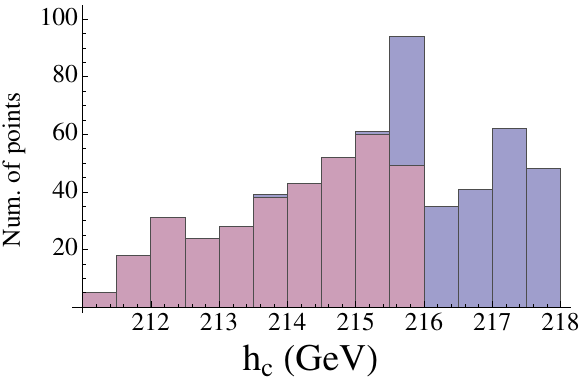} 
\includegraphics[width=0.30\textwidth]{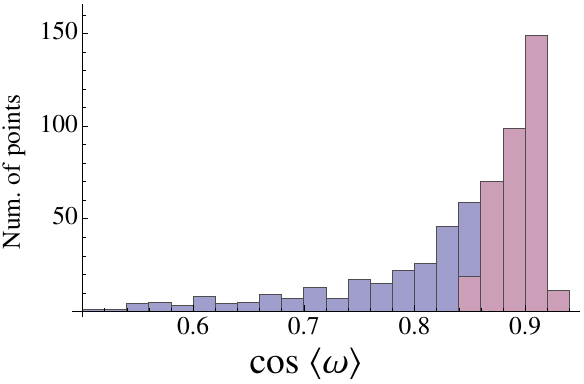} 
\caption{Real singlet Higgs portal model parameters distributions for case (2) that yield a first order EW phase transition. The purple (blue) histograms show the distributions with(out) the constraint $\cos \langle \omega \rangle \geq 0.85 $. For all the model points the Higgs boson mass assumes the correct value. The model parameters have not been constrained using the VCs.}
\label{fig:SMS_c2}
\end{center}
\end{figure}

In Fig.~\ref{fig:SMS_c2_ls} we show how different model parameters relates to $\lambda_s $, whereas the distributions of the model parameters are shown in Fig.~\ref{fig:SMS_c2}.
Same data is used in both figures and all shown points provide $h_c /T_c \geq 1$. In the lower(middle) left panel of Fig.~\ref{fig:SMS_c2_ls} (Fig.~\ref{fig:SMS_c2}) we show how the $m_{0,\Phi} $, i.e. the mass of the new scalar, relates to $\lambda_s $(is distributed). 
The parameter $\cos \langle \omega \rangle$, in the lower right panels of Figs.~\ref{fig:SMS_c2_ls} and \ref{fig:SMS_c2}, determines how much the Higgs couplings change, when compared to the SM values. The SM couplings are reproduced for $\cos \langle \omega \rangle =1$. The purple points/histograms show the distributions once the constraint $\cos \langle \omega \rangle \geq 0.85$ is taken into account. The value $\cos \langle \omega \rangle = 0.85$ is the $2\sigma$ lower bound from the best fit value $\cos \langle \omega \rangle = 0.95$ to the experimental results \cite{Alanne:2014bra}.  
All the points are for the correct Higgs boson mass $\approx 126$ GeV. 
From the top most middle panel of Fig.~\ref{fig:SMS_c2} we note that for the model to be consistent with $\cos \langle \omega \rangle \geq 0.85$ the singlet quartic self coupling $\lambda_{s}$ should assume rather large values. For the smallest values, $\lambda_{s} \sim 5$, the one-loop perturbative approach might still be a reasonable approximation.
{Notice also, that $\Phi$ cannot be DM here because it is unstable. Indeed, $\Phi$ is heavy enough (see lower left panel of Fig.~\ref{fig:SMS_c2_ls}) to decay into two Higgs fields. }

To summarise, the model can provide a strong 1st order EWPT and correct Higgs boson mass. However, to avoid the experimental constraints set for $\cos \langle \omega \rangle$, the singlet quartic self coupling tends to have rather large values $\lambda_{s} \gtrsim 5$. The DM problem cannot  be addressed with this model. 


\subsubsection{Veltman conditions}

The Veltman conditions are Eqs.~\ref{eqVCS} and \ref{eqVCH}.
Once the VC condition, $\lambda_{hs} < 0$, from Eq.~\ref{eqVCS} is satisfied the flat direction condition, Eq.~\ref{flatc0} reads
\begin{equation}
\label{flatc}
\sqrt{\lambda_h \lambda_s} +\lambda_{hs} = 0 \ .
\end{equation}
 This condition agrees with the flatness condition in Eq.~\ref{flatc}.
Using therefore the VCs, Eqs.~\ref{eqVCS} and \ref{eqVCH}, along with the flat direction condition, Eq.~\ref{flatc}, the model parameters get constrained via
\begin{eqnarray}
\label{couplings}
\lambda_{hs}= -\frac{3}{4}\lambda_s \,, \,  \,  \lambda_s=\frac{16}{9} \lambda_h \,, \,  \, \cos^2 \langle \omega \rangle= \frac{4}{7} \,,  \,  \, 
\lambda_h = \frac{9}{56}  ( 8 y^2_t - 3g^2-g'^2 ).
\end{eqnarray}
Implementing these conditions on the Higgs boson mass relation Eq.~\ref{HiggsmassS1}, and using the  measured values of the top Yukawa and the SM gauge couplings, we get  
\begin{eqnarray}
m_{\phi} \approx 95  \, \rm{GeV} \quad \rm{and} \quad m_{\Phi} \approx 541  \, \rm{GeV}.
\end{eqnarray}  Now all the model parameters are fixed and using these values we find that $h_c /T_c \approx 2.2$ indicating the occurrence of  a strong first order EWPT.  Because the mass of the lightest scalar is lower than the one used in the Fig.~\ref{fig:SMS_c2} this model is not present there.   Although the resulting Higgs mass does not agree with experiments to this order in perturbation theory, it would be interesting to investigate higher orders to see if they might alleviate the tension. 

To summarize, once the one-loop VCs are imposed, and holding fixed the top mass to the experimental value, we predict a Higgs boson mass of $m_{\phi} \approx 95$ GeV and strongly first order EWPT as indicated by the $h_c /T_c \approx 2.2$ ratio.  The SM particle couplings to the physical Higgs boson are suppressed by a factor $\cos \langle \omega \rangle$ compared to the SM.  More precisely the model predicts $\cos \langle \omega \rangle \approx 0.76$, which is below the experimental $2\sigma$-limit $\cos \langle \omega \rangle = 0.85$ \cite{Alanne:2014bra}.  

We learnt that the classically conformal singlet Higgs portal model can support a strong 1st order EWPT and provides the correct EW spectrum. This can be viewed as an improvement on the SM.  Further imposing the VC reduces the Higgs mass but strengthen the 1st order EWPT. { Finally, the model can provide at most a subdominant DM component.}

We move now to the next natural step, i.e. adding the simplest new fermionic sector to the theory.


\subsection{SM +  Real Singlet Scalar + Weyl Fermion(s)} 
\label{SMreSwf}

Now we shall consider a model, where the simplest singlet scalar Higgs portal model is extend with a singlet Weyl fermion $\chi$. 
The general non-conformal scalar potential, including the new Weyl fermion $\chi$ interaction term reads
\begin{eqnarray}
\label{ScalarsSMSwf}
V_{0}& =& -\mu^2 H^{\dagger}H  + \lambda_h (H^{\dagger}H)^2 +  m_S^2S^2 + \lambda_{hs}H^{\dagger}HS^2+ \frac{\lambda_{s} }{4} S^4\\
          &+ &\mu_{3}S^3+\mu^3_{1}S+\mu_{m}SH^{\dagger}H + (y_{\chi} S + M_\chi) (\chi \chi+ \overline{\chi}  \overline{\chi} )  \ .  \nonumber
\end{eqnarray}
The EWPT for this kind of models where, however, the new fermion $\chi$ is a Dirac particle has been studied recently in \cite{Fairbairn:2013uta,Li:2014wia,Alanne:2014bra}. {There the particle $\chi$ was associated with dark matter, and it is was found, that the EWPT can be strongly first order and that the model can yield the correct DM density\footnote{Notice, the hard mass term $M_{\chi}\bar{\chi}\chi$ extends the model parameter space, and plays a significant role in the DM studies.}.  {Here too  $\chi$ can be associated with dark matter \cite{Frandsen:2013bfa} and since the new fermion sector has a milder effect than the new boson one on the phase transition we expect  a strong first order phase transition  \cite{Fairbairn:2013uta,Li:2014wia,Alanne:2014bra}.}

We will therefore move to investigate the classically conformal limit of the model.


\subsection{PNC model: Conformal SM + Real Singlet Scalar + Weyl Fermion(s)} 
\label{PNC}

Once we take the classical conformal limit of the Eq.~\ref{ScalarsSMSwf}, the tree level potential reduces to
\begin{eqnarray}
\label{PNC_SM_S_X}
V_{0}& =& \lambda_h (H^{\dagger}H)^2 + \lambda_{hs}H^{\dagger}HS^2+ \frac{\lambda_{s} }{4} S^4+ y_{\chi} S (\chi \chi+ \overline{\chi}  \overline{\chi} ) \ .
\end{eqnarray}
This model resulted to be the most intriguing among the class of {\it the perturbative natural conformality models (PNC)} investigated in \cite{Antipin:2013exa}. The PNC models
attempt to provide the correct Higgs boson mass via the CW mechanism and to alleviate simultaneously the naturality problem by satisfying VC. Obviously these models are much more constrained than either of the two conditions satisfied independently. The above model was, indeed, found to be an ideal candidate PNC model as we shall  briefly review below. It is therefore relevant to see if the model is also capable of leading to a strong 1st order EWPT. 

We also observe that the new classically conformal interactions of  $S$  with the Weyl fermion directly destroys the accidental $Z_2$ symmetry  of the scalar part of the tree potential. 

We shall   consider two cases:  In case (1) only $H$ acquires a non-vanishing VEV.  Where as in case (2) $S$ can also get a non-vanishing VEV.


\subsubsection{Case (1): $\langle s \rangle =$ 0} 
\label{SMSDM_c1}

Here the EWPT investigation is very similar to the one in Sec.~\ref{SMrSc1}. The difference resides in the fact that the singlet {\it dark} fermion affects the singlet scalar thermal mass. The relevant mass terms for the EWPT analysis are given in App.~\ref{AppPNCc1} and \ref{AppHSc1}.

As further constrain we assume the physical Higgs boson mass which allows, as we shall momentarily see, to constrain $\lambda_{hs}$.  We also assumed that the flat direction occurs along the $h$-direction  and thus we  have $\lambda \approx 0$ at the EW scale.  The CW analysis then yields the one-loop mass for the Higgs boson
\begin{eqnarray}
\label{HiggsmassS}
m^2_{\phi} = \frac{3}{8 \pi^2} \left[ \frac{1}{16}(2 g^4+(g^2 +g'^2)^2) + \frac{1}{3}\lambda_{hs}^2- y^4_t \right] v^2_{EW} \,.
\end{eqnarray}
Knowledge of the physical Higgs boson mass value as well as the SM couplings $y_t, g$ and $g'$, yields $\lambda_{hs}\approx 4.84$.  We are left with two free model parameters that we allow to range in 
\begin{eqnarray}
0.01 \leq \lambda_s  \leq 4.5 \quad \rm{and} \quad 0.01  \leq y_{\chi}  \leq 6.5 \, , 
\end{eqnarray}
and search for the parameter space where $h_c /T_c  \geq1$. Our results are reported in Fig.~\ref{fig:PNC_c1}. Similarly to Sec.~\ref{SMrSc1} we observe that the extra contributions to the singlet scalar thermal mass decreases the ratio $h_c /T_c$ and as expected the largest values for $h_c /T_c$ is obtained when $ \lambda_s$ and $y_{\chi}$ are set to zero. From the Fig.~\ref{fig:PNC_c1} we see that the model can provide $h_c /T_c > 1$ values for a large part of the model parameter space, particularly favouring small $\lambda_s$ and $y_{\chi}$ values. Further, using  $\lambda_{hs}\approx 4.84$, the model predicts that the tree level singlet scalar mass is 541 GeV, as it is the case in Sec.~\ref{SMrSc1}.
\begin{figure}
\begin{center}
\includegraphics[width=0.43\textwidth]{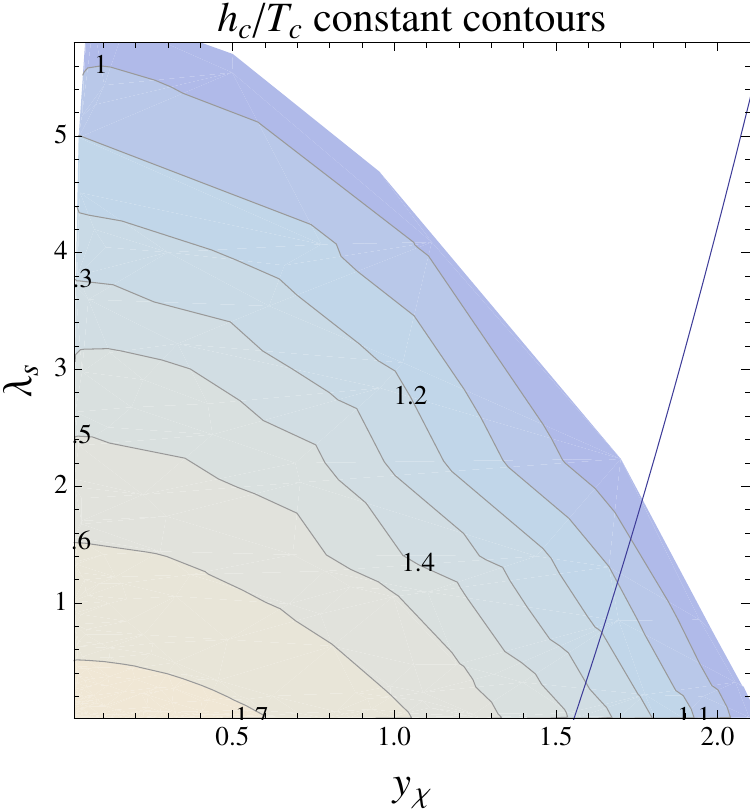} 
\caption{We show the $h_c /T_c$ contour plot as function of the dark Yukawa  coupling $y_\chi$ and the scalar S quartic self coupling $\lambda_{s}$. The portal coupling $\lambda_{hs}$ is fixed to the value $\approx 4.84$, which provides the correct Higgs boson mass and satisfies the VC for the Higgs boson mass term.
On the upward turning blue curve also the VC for the new scalar boson S mass is satisfied.}
\label{fig:PNC_c1}
\end{center}
\end{figure}


\subsubsection{Veltman conditions}

The VCs  read  \cite{Antipin:2013exa}
\begin{eqnarray}
\label{eqVCSDM}
3 \lambda_s+4 \lambda_{hs} - 8 y^2_{\chi} = 0  \,,
\end{eqnarray}
for the singlet, and
\begin{eqnarray}
\label{eqVCHDM}
6 \lambda_h+\frac{9}{4}g^2+\frac{3}{4}g'^2-6y^2_t + \lambda_{hs}=0  \,,
\end{eqnarray}
for the Higgs field.
Compared to the case studied in Sec.~\ref{cSMrS} in Eq.~\ref{eqVCS}, the VC for the singlet has been now modified because of the presence of the dark Yukawa coupling $y_{\chi}$. This allows $\lambda_{hs}$ to be positive here. 
Further, note that the VC for the Higgs boson, Eq.~\ref{eqVCHDM}, gives $\lambda_{hs}\approx 4.84$, which is very surprisingly exactly the value needed to yield the physical Higgs mass in  Eq.~\ref{HiggsmassS}. Using this value of  $\lambda_{hs}\approx 4.84$ in Eq.~\ref{eqVCSDM} the VC returns the relation $ \lambda_s \approx \frac{8}{3} y^2_{\chi}  - 6.45$, shown as the upward turning blue curve in Fig.~\ref{fig:PNC_c1}. Thus, on this curve the Veltman condition for the scalar boson $S$ is satisfied. To collect, the model predictions are
\begin{eqnarray}
\label{resPNC}
m_{\phi}  \approx 126  \, \rm{GeV}, \quad m_{\Phi} \approx 541 \, \rm{GeV}, \quad  \lambda_{hs} \approx  4.84 \quad  \rm{and} \quad   \lambda_s \approx \frac{8}{3} y^2_{\chi}  - 6.45,
\end{eqnarray}
that also give $h_c /T_c \geq 1$ once  $y_{\chi} \lesssim 1.7$.

To summarize once the VCs are imposed, the model becomes very constrained and thus highly predictive. Remarkably it yields the correct mass for the Higgs boson and predicts the mass for the singlet scalar while satisfying the VCs \cite{Antipin:2013exa}. In addition, the model supports a first order EWPT.  Although the value of $\lambda_{hs} \approx 4.84$ is not very small it is still not unreasonably large for a perturbative study to hold \cite{Tamarit:2014dua}. Nevertheless higher order corrections would be interesting to investigate. Further, the VC for the singlet can be satisfied for very small couplings with the limiting case of $\lambda_s \approx 0$ once $y_{\chi} \approx 1.55$ providing also the largest ratio $h_c /T_c \approx 1.4$.

\subsubsection{Dark aspects}

Here we briefly discuss aspects related to the new Weyl fermion $\chi$. As shown in the previous section,
once we take the classical conformality and the VC to guide the model building,   
the dark Yukawa coupling $y_{\chi}$ gets connected to $\lambda_s$ in Eq.~\ref{resPNC}. 
If we furthermore ask for the largest possible $h_c /T_c$ ratio, the model predicts $y_{\chi} \approx 1.55$. Because, however,  $S$ does not acquire a VEV this Weyl fermion $\chi$ remains massless at zero temperature. Unless we further extend the model $\chi$ would behave as a sterile light neutrino.  If we allow the introduction of a small conformal breaking by intruding a hard mass term (or yet another scalar field acquiring a VEV) for $\chi$ then $\chi$ could be considered as a DM candidate \cite{Frandsen:2013bfa}. We will not speculate further on this point and move to the next natural case in which also $S$ can receive a non-vanishing VEV via the CW mechanism. In this case $\chi$ will acquire a tree level mass term while abiding the classical conformality requirement.

\subsubsection{Case (2): $\langle s \rangle \neq$ 0} 
\label{PNCc2}

We will now consider the general case in which also the singlet $S$ acquires a non-vanishing VEV. For this case we follow the steps in Sec.~\ref{SMS_c2}.
To have a non-trivial flat direction and EW symmetry breaking via CW the model parameters should satisfy the relations in Eq.~\ref{eqflatnontri} and Eq.~\ref{flatc0}. For potential stability reasons the scalar couplings should further satisfy the relation $\lambda_{h,s} \geq 0$.  

The mostly singlet state has a tree level mass $m^2_{0,\Phi} =2(\lambda_h-\lambda_{hs}) v^2_{EW}$ as in Sec.~\ref{SMS_c2}. However, because the $S$ field receives a VEV, and $H$ and $S$ mix, the 
one-loop Higgs mass gets modified to,  
\begin{eqnarray}
\label{HiggsmassSc2}
m^2_{\phi} = \frac{\cos^2  \langle \omega \rangle}{8 \pi^2} \left[ \frac{1}{16}(6 g^4+3(g^2 +g'^2)^2) +4 (\lambda_h-\lambda_{hs})^2- \frac{12}{4}y^4_t - 32 y^4_{\chi} \tan^4  \langle \omega \rangle \right] v^2_{EW} \,.
\end{eqnarray}
This means that also $\chi$ contributes to the Higgs boson mass, even though the effect is suppressed by the mixing.
Using the flatness condition, Eq.~\ref{flatc0}, and the Higgs mass relation, Eq.~\ref{HiggsmassSc2}, we may relate and constrain the model parameters $\lambda_h, \lambda_s, \lambda^2_{hs}$ and $y_{\chi}$ as in Sec.~\ref{SMS_c2}. However, compared to Sec.~\ref{SMS_c2}, here we have one more parameter $y_{\chi}$, and thus we made the scan over two parameters within the ranges 
\begin{eqnarray}
0.01 \le y_{\chi} \le 5 \quad \rm{ and}   \quad 0.01 \le \lambda_s \le 4 \pi.
\end{eqnarray}
We used flat prior distributions with 2000 points in the scan.
In Fig.~\ref{fig:PNC_c2} we show $h_c /T_c \geq 1$ constant contours in $(y_{\chi},\lambda_{s})-$plane. 
 \begin{figure}
\begin{center}
\includegraphics[width=0.43\textwidth]{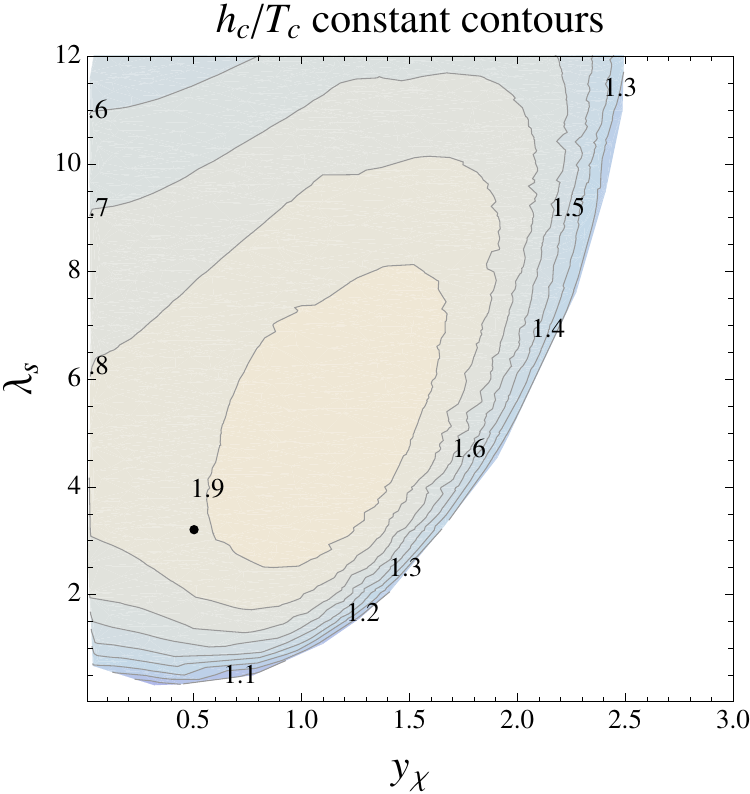} 
\caption{We provide the $h_c /T_c$ contour plot in the ($y_{\chi},\lambda_{s}$)-plane {for the PNC model of case (2)}.  The experimentally observed value of the Higgs boson mass $\approx 126$ GeV is enforced. The black dot shows the point where the model satisfies also the VCs and it leads to the values $\cos \langle \omega \rangle  \approx 0.79$, $m_{\chi} =191$ GeV and $m_{\Phi} = 604$ GeV.
}
\label{fig:PNC_c2}
\end{center}
\end{figure}
We observe that it is possible to achieve the phenomenologically desirable correct value of the Higgs mass for a very wide range of the parameter space along with a 1st order EWPT. 

\begin{figure}[t]
\begin{center}
\includegraphics[width=0.30\textwidth]{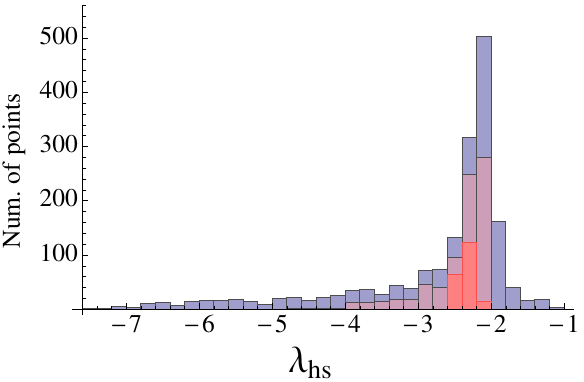} 
\includegraphics[width=0.30\textwidth]{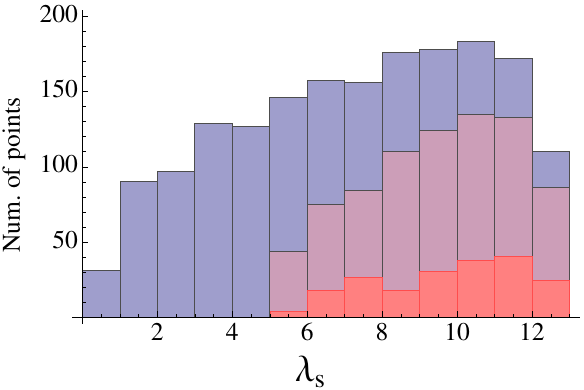} 
\includegraphics[width=0.30\textwidth]{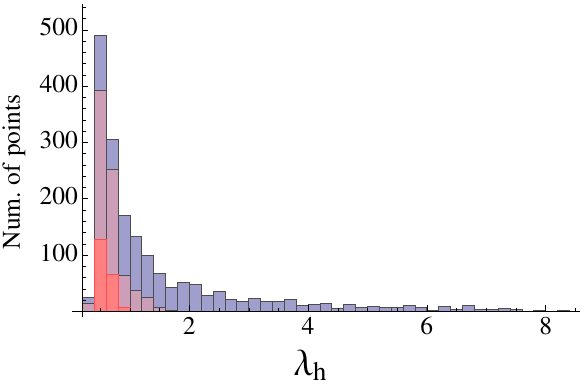} 
\includegraphics[width=0.30\textwidth]{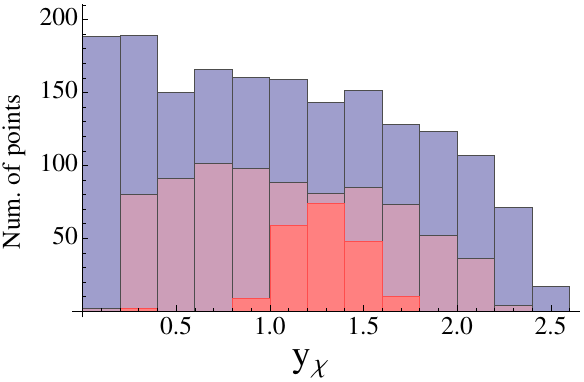} 
\includegraphics[width=0.30\textwidth]{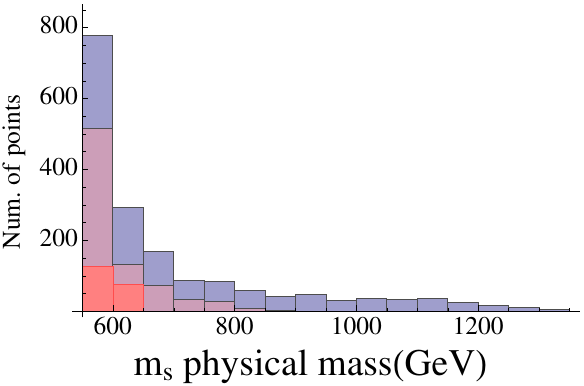} 
\includegraphics[width=0.30\textwidth]{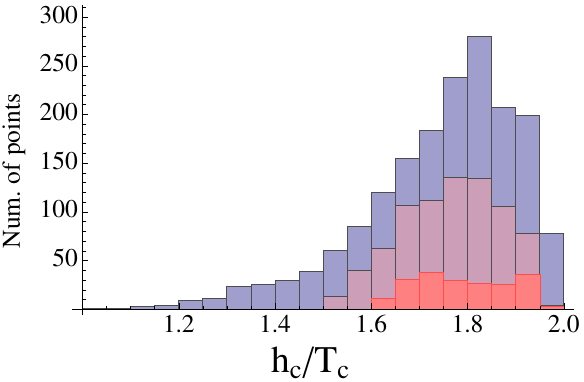} 
\includegraphics[width=0.30\textwidth]{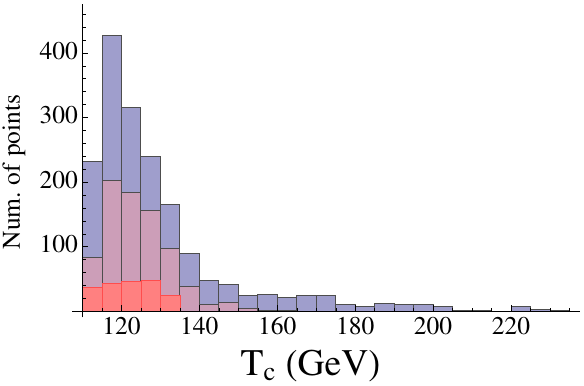} 
\includegraphics[width=0.30\textwidth]{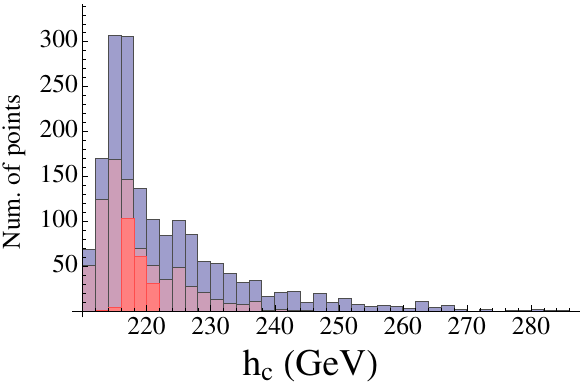} 
\includegraphics[width=0.30\textwidth]{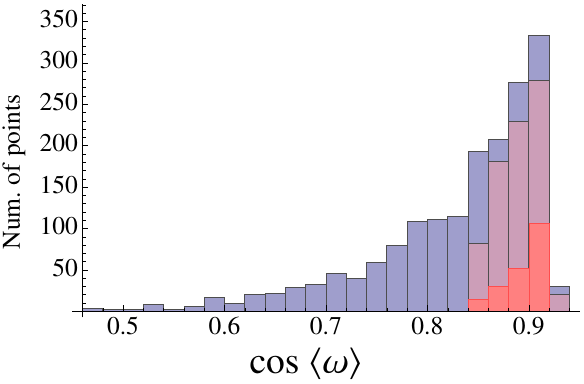} 
\caption{Distributions of the PNC model parameters in case (2) producing first order EW phase transition. The purple (blue) histograms show the distributions with(out) the constraints $\cos \langle \omega \rangle \geq 0.85 $ and $\rm{BR}_{\phi \rightarrow \chi \chi} < 0.17$. The red points, providing a subdominant DM component, survive also the LUX \cite{Akerib:2013tjd} constraint.}
\label{fig:PNC_c2H}
\end{center}
\end{figure}
 
\subsubsection{Veltman conditions}

The VCs for this model are Eqs.~\ref{eqVCSDM} and \ref{eqVCHDM}.  Using these with the flat direction condition, Eq.~\ref{flatc0}, and the Higgs mass relation, Eq.~\ref{HiggsmassS} and using the measured values for the couplings $g, g'$ and $y_t$, we can solve for all the four free parameters of the model. Keeping only the solutions yielding positive real valued physical quantities, the model parameters are constrained to be
{
\begin{eqnarray}
\label{parmPNC}
\lambda_h \approx 1.12 ,\quad \lambda_{hs} \approx -1.90,\quad \lambda_{s} \approx 3.20,\quad y_{\chi} \approx 0.50,
\end{eqnarray}
}
yielding the following physical predictions 
\begin{eqnarray}
\label{parmPNCphys}
m_{\phi} \approx 126 \, \rm{GeV}, \quad m_{\Phi} \approx  604 \, \rm{GeV},\quad m_{\chi} \approx 191 \, \rm{GeV} ,\quad \cos \langle \omega \rangle  \approx 0.79.
\end{eqnarray}
Further, using the parameter values Eq.~\ref{parmPNC} for the EWPT study we find $h_c /T_c \approx 1.88$ indicating a sufficiently strong first order phase transition.

Let's pause and reflect on the results. Notably, all the couplings are fixed and relatively small, thus the perturbative analysis for the phase transition is robust. The model yields the correct Higgs boson mass values and predicts one extra singlet scalar and Weyl fermion with masses $m_{\Phi} \approx  604 $ GeV and  $m_{\chi} \approx 191 $  GeV respectively. 
The Higgs coupling factor $\cos \langle \omega \rangle  \approx 0.79$ is slightly below the experimental 2$\sigma$-limit and higher loop corrections to the Higgs mass relation and VCs could further alleviate the tension. Since the $H$ and $S$ fields mix we have that $\chi$ couples to the Higgs via the portal coupling. Thus, if $\chi$ is assumed to be a DM candidate, because $y_{\chi}$ is not that small and furthermore $m_{\chi}$ is naturally predicted to be in the {\it classic} WIMP mass region, it could be probed via DM direct detection experiments. 
On the other hand $\chi$ could equally be associated with a massive right handed sterile neutrino, which could then mass mix with the SM neutrinos. Of course, as we have shown the VCs can be relaxed an a much more general parameter space is allowed consistent with a 1st order EWPT and phenomenological constraints. 

It is therefore relevant to discuss next the intriguing dark aspects of this model.

\subsubsection{Dark aspects and constraints}

We discuss now shortly about the constraints and aspects related to two different $\chi$ scenarios. The one in which $\chi$ is the DM candidate and the other in which $\chi$ is a right handed neutrino. We first consider the most evident constraints for the former case, which we call scenario A. These constraints can be partially used also for the latter case, called scenario B.  
 
\emph{\underline {Scenario A.}}
Here we star by assuming, that $\chi$ is a DM particle stabilised by a new discrete symmetry e.g.~$Z_2$. We compute its annihilation cross sections into SM fermions, gauge and Higgs bosons as well as into the new scalar $\Phi$.  We evaluate the relic density using the classic approximate method to check if it can account for the full cosmic DM relic density $\Omega _{\rm DM}$.  The details can be found in the App.~\ref{AppAnn}. However, we consider a wider parameter space where $\chi$ can also be a subdominant component quantified via $f_{rel} = \frac{\Omega _{\chi }h^2}{\Omega _{\rm{DM}}h^2}$. Here $\Omega _{\chi }h^2$ is the $\chi$  density parameter (times the dimensionless Hubble parameter squared) and ${\Omega _{\rm DM}h^2}=0.1188$ is the DM one consistent with the latest observations \cite{Ade:2015xua}.  
Using the DM direct detection constraints and the Higgs invisible decay branching ratio bound we can now limit the parameter space of the model.  

We first consider the more general model scenario, where the VCs are not assumed. 
The spin independent  $\chi$-nucleon cross section for the model is\footnote{The minus sign in the last squared term in Eg.~\ref{SI} follows directly from the real orthogonal mixing matrix which diagonalizes the (h,s) mass matrix. In our case one of the mass eigenvalues is positive and the other is zero, i.e.~the Higgs-like state has zero mass at tree level. 
}
\begin{eqnarray}
\label{SI}
\sigma_{\rm SI} = \frac{4 f^2 \mu^2}{\pi } y^2_{\chi} \sin^2 \langle \omega \rangle \cos^2 \langle \omega \rangle   \left( \frac{1}{m_{\phi}^2}- \frac{1}{m_{\Phi}^2}\right)^2, 
\end{eqnarray}
where $\mu$ is the $\chi$-nucleon reduced mass and $f$ is the effective Higgs-nucleon coupling, for which we used the value $f =0.345$ \cite{Cline:2013gha}.
{ 
To ease the comparison with the DM direct detection constraints  also for the subdominant case we define the following effective scattering cross section (see e.g.~\cite{Cline:2012hg,Cline:2013gha,Kainulainen:2015raa})
\begin{eqnarray}
\label{SIeff}
\sigma^{\rm eff}_{\rm SI} \equiv \sigma_{\rm SI} f_{\rm rel} < \sigma^{\rm exp}_{\rm SI},
\end{eqnarray}
where  $\sigma^{\rm exp}_{\rm SI}$ is the experimental cross section bound.  
Naturally, the usual constraint for the cross section is obtained once $f_{\rm rel} =1$.
To constrain the model we use the recent LUX results \cite{Akerib:2013tjd} for $\sigma^{\rm exp}_{\rm SI}$.

Let us now explain why it is convenient to use $\sigma^{\rm eff}_{\rm SI} $.
Under the expectation that DM clusters similarly in all cosmic structures irrespectively on its sub- or dominant nature, the factor $f_{\rm rel}$ can be used to properly scale the local $\chi$ density via the relation $\rho^{\rm loc}_{\chi}= f_{\rm rel} \, \rho^{\rm loc}_{\rm DM}$, where $\rho^{\rm loc}_{\rm DM}\approx 0.3 \, \rm{GeV / cm^3}$. This affects the interpretation of the DM direct detection bounds that depends on the assumed local DM density.  Indeed, as the local density for the subdominant DM is smaller than the usual $\rho^{\rm loc}_{\rm DM}$, the scattering cross section $\sigma_{\rm SI}$ in Eq.~\ref{SI}, can be larger than $\sigma^{\rm exp}_{\rm SI}$ obtained using $f_{\rm rel}=1$.
 Thus, better quantity to constrain is the effective scattering cross section $\sigma^{\rm eff}_{\rm SI} $, defined in Eq.~\ref{SIeff}, which takes properly into account the local DM density scaling for each type DM component.

In the Fig.~\ref{fig:PNC_c2_SI} we show the scatter plot of the model cross section $\sigma^{\rm eff}_{\rm SI}$ and the LUX 90$\%$ CL limit \cite{Akerib:2013tjd}. 
Colorcoding indicates the relative dark matter density with red giving the full DM density where as blue only a percent level fraction. Grey points yield too much dark matter i.e.~$f_{\rm rel} >1$ and thus are excluded.   Triangles pass the other experimental constraints, like the invisible Higgs decay $\rm{BR}_{\phi \rightarrow \chi \chi} < 0.17$ constraint discussed below, and also the experimental $2\sigma$-limit $\cos \langle \omega \rangle \ge 0.85$ mentioned previously. Dots do not pass these extra constraints.}

In the case $\chi$ is light, such that $2m_{\chi} \leq m_{\phi}$, it can open up an invisible Higgs decay channel. In this case the invisible Higgs decay width to $2\chi$ is
\begin{eqnarray}
\label{Hdecay}
\Gamma_{\phi \rightarrow \chi \chi} = \frac{m_{\phi}}{4 \pi } y^2_{\chi} \sin^2 \langle \omega \rangle  \left( 1- \frac{4 m^2_{\chi}}{m_{\phi}^2}\right)^{3/2}, 
\end{eqnarray}
where $m_{\phi}$ is the physical Higgs mass and $m_{\chi} = 2 y_{\chi} \tan \langle \omega \rangle v_{EW}$. The invisible Higgs decay branching fraction in this case reads 
\begin{eqnarray}
\rm{BR}_{\phi \rightarrow \chi \chi} = \frac{\Gamma_{\phi \rightarrow \chi \chi}}{\Gamma_{SM} + \Gamma_{\phi \rightarrow \chi \chi}}, 
\end{eqnarray}
where $\Gamma_{\rm SM}$ is the Higgs decay width in the SM. This branching fraction is constrained by the LHC data, and for the limit we used $\rm{BR}_{\phi \rightarrow \chi \chi} < 0.17$, where the 95$\%$ CL limit $\rm{BR}_{\rm inv.} < 0.17$ was obtained from \cite{Giardino:2013bma}. 
\begin{figure}
\begin{center}
\includegraphics[width=0.55\textwidth]{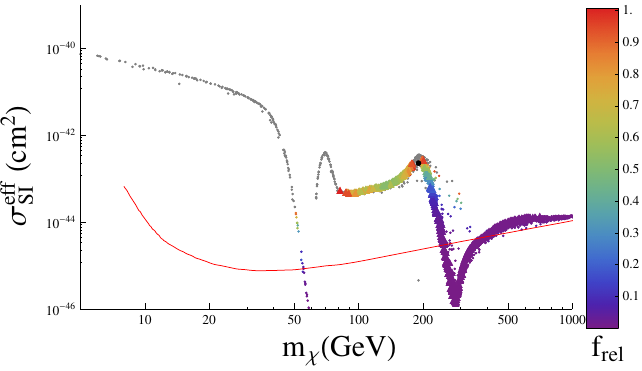} 
\caption{The model effective spin independent $\chi$-nucleon cross section $\sigma^{\rm eff}_{\rm SI}$ as a function of DM mass $m_{\chi}$. Colorcoding depicts the relative DM density $f_{\rm  rel}$ for each point. 
Grey points have $f_{\rm rel}>1$ and are thus excluded. Large triangles pass the invisible Higgs decay branching fraction limit and the experimental $2\sigma$-limit $\cos \langle \omega \rangle \ge 0.85$, whereas small dots do not. The black dot indicates the parameters space point satisfying the VCs.
The red solid curve is the LUX 90$\%$ CL limit \cite{Akerib:2013tjd} above which the model points are excluded. 
The model data match the ones used in Fig.~\ref{fig:PNC_c2H} and furthermore for all the shown points $h_c /T_c \geq 1$.
}
\label{fig:PNC_c2_SI}
\end{center}
\end{figure}

{
Let us now check the scenario where the VCs for the model are satisfied. In this case all the model parameters are fixed and given in Eqs.~\ref{parmPNC} and \ref{parmPNCphys}.
Using these values for the relic density analysis we get that $\Omega _{\chi }h^2 \approx 0.164$ providing $f_{\rm  rel} \approx 1.377$, which is surprisingly close to the full cosmic DM density.  However, once we calculate the scattering cross section, using the values from Eqs.~\ref{parmPNC} and \ref{parmPNCphys} in Eq.~\ref{SI}, we get that $\sigma_{\rm SI} \approx 1.56 \times 10^{-43}$ and comparing with LUX limit \cite{Akerib:2013tjd}, $\sigma^{\rm LUX}_{\rm SI}(m_{\rm DM} \approx 190 \, \,\rm{GeV}) \approx 2.2 \times 10^{-45} \,\, \rm{cm}^2$, we notice that the model point, indicated as a black dot in Fig.~\ref{fig:PNC_c2_SI}, is ruled out.
In fact, this large $\sigma_{\rm SI}$ cross section is excluded also by several other direct DM detection experiments.}
  
 {To summarize, the model can yield a phenomenologically viable subdominant DM component, with $\lesssim$ few $\%$ of the total observed DM density without the VCs.}
 
\emph{\underline{Scenario B.}} If we associate $\chi$, not with DM but simply with a massive right handed neutrino we have the same constraints except the DM direct detection limit $\sigma_{\rm SI}$ limit. Thus all the purple points in Figs.~\ref{fig:PNC_c2H} and \ref{fig:PNC_c2_SI} are phenomenologically acceptable in this scenario. 
However, once VCs are imposed the model is in tension with the experiments because $\cos \langle \omega \rangle  \approx 0.79$. 
 
To summarise, the model allows for the observed value of the Higgs boson mass and first order EWPT. It also predicts a new massive scalar and massive Majorana particle.  The VCs are still very constraining and although the correct Higgs boson mass is still viable together with strong EWPT, the Higgs couplings to the SM fields are suppressed by a factor $\cos \langle \omega \rangle  \approx 0.79$ and deviating by more than two sigma from the LHC lower bound result $\cos \langle \omega \rangle \ge 0.85$.  

We move now to complex scalar extensions of the SM. 

\section{(Non)conformal Extension with A Complex Scalar}
\label{SMCS}

In this section we consider the EWPT in minimal complex scalar extensions of the SM. We start by briefly discussing the EW singlet complex scalar Higgs portal models, and  end with the "SE$\chi$y" model, where the new scalar is not a SM singlet.

\subsection{SM + Complex Singlet Scalar}

The tree level scalar potential of the complex scalar Higgs portal model reads
\begin{eqnarray}
\label{Scalars}
V_{0} = -\mu^2 H^{\dagger}H  + \lambda_h (H^{\dagger}H)^2
 +  m_S^2S^{\dagger}S + \lambda_{hs}H^{\dagger}HS^{\dagger}S + \lambda_{s} (SS^{\dagger})^2\ .
\end{eqnarray}
Phase transitions in this type of models have been investigated extensively in the past e.g.~\cite{Anderson:1991zb,Espinosa:1993bs,Tamarit:2014dua} \footnote{EWPT and DM aspects of more general potential than Eq.~(\ref{Scalars}), including softly U(1)-symmetry breaking terms, are studied in \cite{Barger:2008jx,Gonderinger:2012rd,Jiang:2015cwa}.}. 
For this reason we shall not make a full EWPT analysis here but concentrate instead on the classically conformal limit, and on extensions, that have not been considered in the literature.   
 
\subsubsection{Conformal SM + Complex Singlet Scalar}

The classically conformal limit of the potential in Eq.~\ref{Scalars} is
\begin{eqnarray}
\label{conScalars}
V_{0} =  \lambda_h (H^{\dagger}H)^2 + \lambda_{hs}H^{\dagger}HS^{\dagger}S + \lambda_{s} (SS^{\dagger})^2\ .
\end{eqnarray}
 If only the Higgs field direction is allowed to undergo a radiative CW symmetry breaking  scenario  the EWPT analysis is similar to the one in Sec.~\ref{SMrSc1}. 
Now instead of one real scalar we have two, and adding more real scalars (up to 12) leads to the case studied in \cite{Espinosa:2007qk,Espinosa:2008kw}.
These results are verified and further improved, for the complex $S$ scalars, via a 2-loop RG improved EWPT analysis in \cite{Tamarit:2014dua}.
There it is demonstrated, that a strong 1st order phase transition can be obtained reliably from perturbative analysis even with large portal coupling values.
Since this model has been extensively studied elsewhere, we will not repeat here the EWPT study.  { Finally, adopting the results of \cite{Cline:2013gha}, we note that the model can provide a subdominant DM component.}

\subsection{SM + Complex Singlet Scalar + Fermions}

If new singlet Weyl fermions $\chi_i$ are added to the model, one arrives to the Majoron-like model \cite{Chikashige:1980ui}. The tree-potential is
\begin{eqnarray}
V_{0} &=& -\mu^2 H^{\dagger}H  + \lambda_h (H^{\dagger}H)^2
 +  m_S^2S^{\dagger}S + \lambda_{hs}H^{\dagger}HS^{\dagger}S + \lambda_{s} (SS^{\dagger})^2  \nonumber \\ 
 &+&  \frac{1}{2} \sum_i y_{\chi_i} S \chi_i \chi_i + h.c.
\end{eqnarray}
The U(1) symmetry in the $(S,\chi)$-sector can be associated with the lepton number which can be  spontaneously broken if $S$ receives a VEV. The Goldstone boson related to this broken symmetry is known as the Majoron. Furthermore when  $S$ acquires a VEV, the $y_{\chi_i} S \chi_i \chi_i + h.c.$ term produces a Majorana mass term for $\chi_i$ which can be identified with the right handed neutrino.  In this scenario the standard model neutrinos acquire a mass via a seesaw mechanism. The EWPT in the Majoron model has been studied in \cite{Kondo:1991jz,Enqvist:1992va,Sei:1992np} and more recently in \cite{Cline:2009sn} including the LHC phenomenological reach. After the Higgs boson discovery at the LHC, the Majoron model parameters space, studied in \cite{Cline:2009sn}, are severely constrained. 

\subsubsection{Conformal SM + Complex Singlet Scalar + Fermions}

The EWPT phase transition in a model, closely related to the classically conformal Majoron model, has been recently studied in \cite{Farzinnia:2013pga,Farzinnia:2014xia,Farzinnia:2014yqa}. When requesting for 1st order phase transition, the model becomes rather constrained and predictive. For example, the mass of the new pseudo scalar "Majoron", which is also a subdominant dark matter candidate, is constrained to be in the range $\sim$ 1 - 2 TeV and the mass of the new CP-even scalar is  $\sim $ 100 - 200 GeV. 
 {The next LHC run could provide useful to investigate the model predictions \cite{Farzinnia:2014yqa}.}

\subsection{SE$\chi$y model: SM + Complex Scalar + Fermions}
\label{sxymodel}

Now we will consider the EWPT in a model, which was originally build for DM phenomenology.
This "SE$\chi$y" model was introduced in \cite{Dissauer:2012xa}, and features a new complex scalar field $S$ charged under U(1) hypercharge, a vector like heavy electron $E$ and a neutral fermion $\chi$, which is the dark matter candidate. 
The complex scalar has tree level interactions with the standard model Higgs field (Higgs portal type), whereas the dark matter particle has only loop-induced interactions with the standard model particles. The original motivation for this model was to construct an UV completion of the effective magnetic moment dark matter model \cite{DelNobile:2012tx} able to alleviate, at the time, the tension between the results of different DM direct detection experiments.
The model phenomenology, including the dark matter and collider aspects, has been further studied in \cite{Dissauer:2012xa,Frandsen:2013bfa}. The possibility to secure the SM vacuum stability in the SE$\chi$y model context has been studied in \cite{Antipin:2013bya}.
Here we further study the model parameter space allowing for a strong first order EWPT. 

The difference with respect to more traditional complex Higgs portal models is that in our case the new state carries SM interactions\footnote{A study, considering the EWPT and collider aspects, for similar type model, including charged scalar, has been recently performed in \cite{Katz:2014bha}.}. 

The SE$\chi$y model Lagrangian reads \cite{Dissauer:2012xa}:
\begin{eqnarray}
\label{sxy}
    \mathcal{L}_{S\overline{E}\chi {y_{\chi}}}& =&  \mathcal{L}_{SM} + \bar{\chi}i\slashed{\partial}\chi - m_\chi\bar{\chi}\chi + \overline{E}i\slashed{D}E - m_E\overline{E}E - (S \overline{E}{\chi}  y_{\chi} + \text{h.c.} )
    \nonumber \\
& +& D_\mu S^{\dagger} D^\mu S - m_S^2S^{\dagger}S-\lambda_{hs}H^{\dagger}HS^{\dagger}S - \lambda_{s} (SS^{\dagger})^2\ ,
\end{eqnarray}
where $ D^\mu=\partial^\mu-ie\frac{s_w}{c_w}Z^{\mu}+ieA^\mu$, $s_w$ and $c_w$ refer to the sine and cosine of the Weinberg angle. 
We assume that the couplings $y_{\chi}, \lambda_{hs} $ and $ \lambda_{s}$ are real.
   $\mathcal{L}_{SM} $ stands for all the SM terms. The $S\bar{E} \chi y_{\chi}$ operator in Eq.~\ref{sxy} induces interactions between the dark matter candidate $\chi$ and the SM fields via loop-processes and gives the name to the model. Specifically, quantum corrections induce an effective magnetic dipole moment $\lambda_{\chi}(q^2)$ for the DM state. See e.g.~Eq.~3 in \cite{Dissauer:2012xa} for more details.

The scalar potential of the theory is identical to the one in Eq.~(\ref{Scalars}) but $S$ is charged under the SM $U(1)$ interactions. Depending on the values of the model parameters the model can feature different symmetry breaking patterns and different type EWPTs.

If $S$ acquires a VEV at zero temperature, the $U(1)$ (hypercharge) symmetry breaks and the photon is massive. Thus we require the tree level potential to yield a vacuum structure of the type $(\langle H \rangle,\langle S \rangle)=(v_{EW},0)$.  Additionally we also require potential stability that leads to two different cases.   In case (1) the $S$ scalar does not generate a VEV at any temperature i.e.~$\langle S \rangle  = 0$, while in case (2) $S$ is allowed to acquire a VEV only at some finite temperature i.e., $\langle s(T) \rangle \neq 0$ at some $T \neq 0$.


\subsubsection{Case (1): $\langle s(T_c) \rangle =$ 0, SM like 1-loop scenario} 
\label{sxyc1}

The tree level potential parameters satisfy the relations
\begin{eqnarray}
\{ \mu^2, \lambda, m^2_S, \lambda_{s}, \lambda_{hs} \} > 0 \ ,
 \end{eqnarray}
so that at the tree level we have $\langle H \rangle = (0,v_{EW}/\sqrt{2})^T$ and  $S$ does not generate a VEV at any temperature i.e.~$\langle S \rangle  = 0$ and we call this case: The SM-like 1-loop scenario.
 The relevant mass terms used in the EWPT analysis are reported in App.~\ref{sexc1} and we scanned  the parameters within the ranges
\begin{eqnarray}
\label{sxyc1P}
0.01 \leq \lambda_{hs} \leq 8\,,  \quad 0.01 \leq \lambda_{s} \leq 4\,, \quad 0.01 \leq y_{\chi} \leq 4 \ .
 \end{eqnarray}
{ Again we used flat prior distributions with 2000 points in the scan.}
\begin{figure}[t]
\begin{center}
\includegraphics[width=0.30\textwidth]{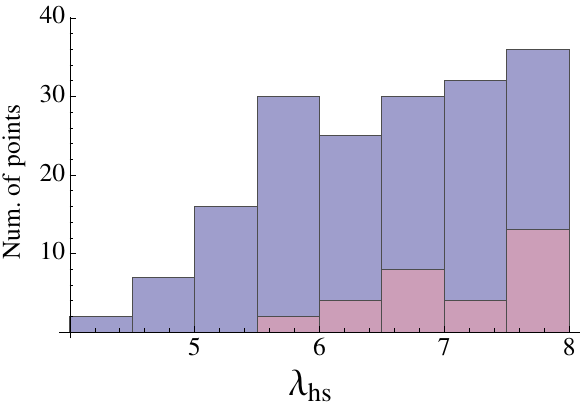} 
\includegraphics[width=0.30\textwidth]{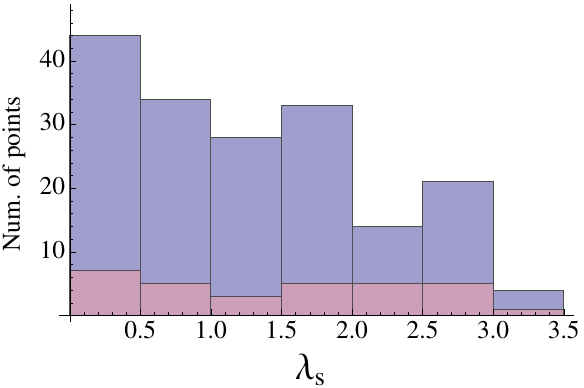} 
\includegraphics[width=0.30\textwidth]{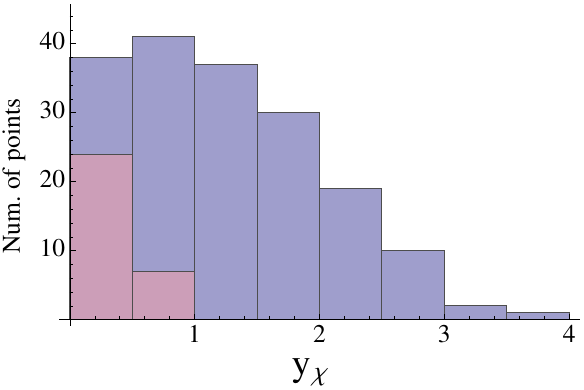} 
\includegraphics[width=0.30\textwidth]{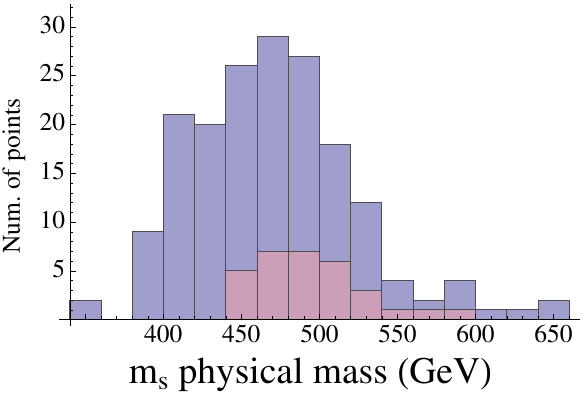} 
\includegraphics[width=0.30\textwidth]{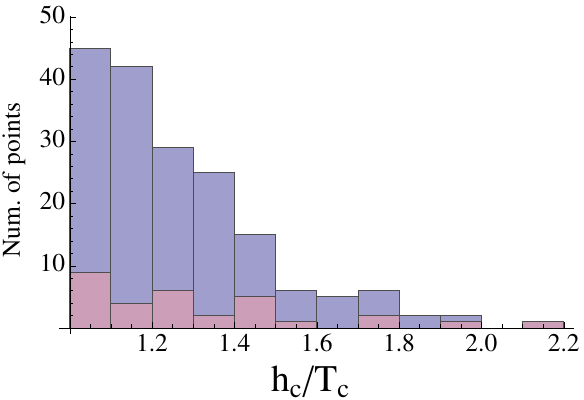} 
\includegraphics[width=0.30\textwidth]{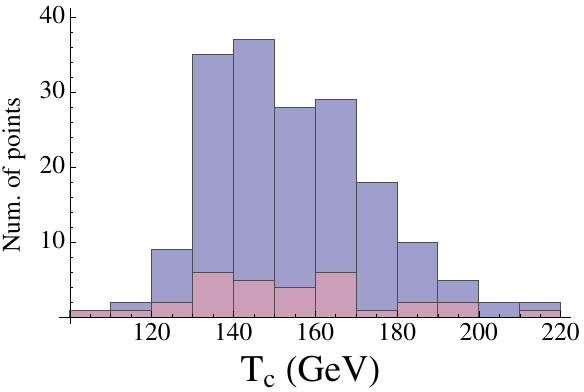} 
\includegraphics[width=0.30\textwidth]{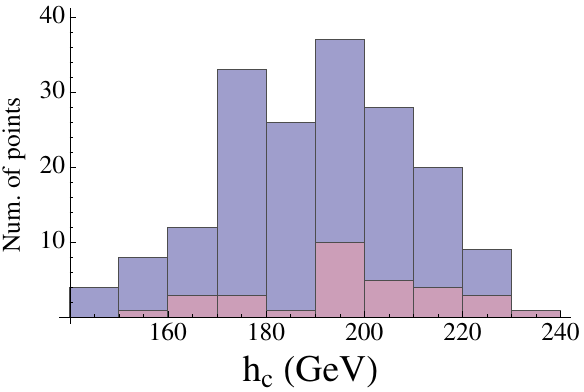} 
\includegraphics[width=0.30\textwidth]{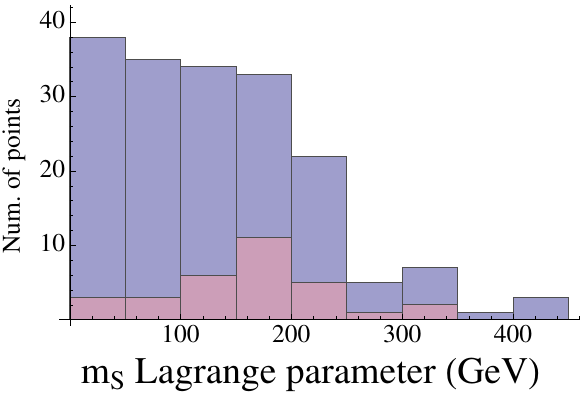} 
\caption{We show the results of the parameter scan for the SE$\chi$y model in case (1) producing a first order EWPT.  The purple (blue) histograms shows the distributions of the points with (without) the constraints $m_s >  450$ GeV and $y_{\chi} < 0.6$. We have not imposed the VCs. }
\label{fig:sxy_c1}
\end{center}
\end{figure}
 In Fig.~\ref{fig:sxy_c1},  we show the results of the scan and show that the model supports a first order phase transition occurring for relatively light values of the new charged scalar that can be constrained by LHC data, as we shall discuss below. Consistently with the findings in the previous sections $\lambda_{hs}$ tends to be quite large.  

\subsubsection{Dark aspects and constraints}
\label{sxyDM}

The DM phenomenology has been studied in \cite{Dissauer:2012xa,Frandsen:2013bfa}. However in \cite{Frandsen:2013bfa} the value  $y_{\chi}=2$, motivated by the VCs, was used to constrain the model. Relaxing the VCs on $y_{\chi}$, as we shall see, will improve on the model phenomenological viability.  

We can start by noticing that the LUX experiment directly constrains the effective magnetic moment coupling, $\lambda_{\chi}(0) \simeq \frac{e y_{\chi}^2}{32 \pi^2 m_s}$.  Clearly, allowing for smaller values of $y_{\chi}$ than in \cite{Frandsen:2013bfa} reduces also the lower limit on  $m_s$ limit. Smaller values of $m_s$ are welcomed since they are expected to lead to a stronger first order EWPT.  We also assume that $m_E < m_{\chi} < m_s$ which allows for a phenomenologically viable DM candidate. 

Following \cite{Dissauer:2012xa} we allow E to have small mixing with SM leptons (tau) such that $S$ can  decay into $\chi$ and $E$, and $E$ decays further in the SM fields. Assuming the collider limit for $ m_E \gtrsim 393$ GeV \cite{Dissauer:2012xa} we take $m_{\chi} \gtrsim 400$ GeV. The dominant DM annihilation channel is $\bar{\chi} \chi \rightarrow \bar{E} {E} $ in the $ t-channel$ which occurs via the $S-$exchange, see Eq.~10 of Ref.~\cite{Frandsen:2013bfa}
\begin{eqnarray}
\label{sigann} 
\langle \sigma^{S}_{\bar{\chi} \chi} v_{rel} \rangle = \frac{y^4 (m^2_E + m^2_{\chi} )}{8 \pi (m^2_E - m^2_{\chi} -m^2_s)^2} \sqrt{1- \frac{m^2_E }{m^2_{\chi}}}+ {\cal{O}}(v^2)
 \end{eqnarray}
For thermal (Dirac) DM the annihilation cross section is $\langle \sigma  v_{rel} \rangle \approx 4.4 \times 10^{-26} \frac{\rm cm^3}{s}$ for $m_{\rm DM} \gtrsim 15$ GeV\cite{Steigman:2012nb}. Using this cross section with Eq.~\ref{sigann} and assuming, for example  $m_s = 500$ GeV, $m_{\chi} =420$ GeV and $ m_E= 400$ GeV, we derive $y_{\chi} \approx 0.425$. Such small values of $y_{\chi}$  and relatively light S, i.e. $m_s \simeq 500$~GeV, are now compatible with LUX that now allows scalar masses larger than $m_s \approx 90$ GeV. A more refined analysis that takes into account also the correct thermal DM  relic density leads to the constraints  $m_s >  450$ GeV and $y_{\chi} < 0.6$. The purple histograms in Fig.~\ref{fig:sxy_c1} show the parameter space passing these constraints.  The heavy electron mass does not affect the EWPT analysis unless it is heavy enough to start becoming Boltzmann suppressed. {If this is the case, the $y_{\chi}$ dependence is suppressed in the $S$ boson thermal mass and thus the transition becomes very weakly dependent on $y_{\chi}$.}

\subsubsection{Veltman conditions}

The VC for the new charged scalar  reads \cite{Antipin:2013exa}
\begin{eqnarray}
\label{eqVCSsxy}
4 \lambda_s+2 \lambda_{hs} +3g'^2 - 2y^2_{\chi} = 0  \,,
\end{eqnarray}
and for the Higgs field  
\begin{eqnarray}
\label{eqVCHsxy}
6 \lambda_h+\frac{9}{4}g^2+\frac{3}{4}g'^2-6y^2_t + \lambda_{hs}=0  \,.
\end{eqnarray}
Using the known SM coupling values in Eq.~\ref{eqVCHsxy} we get that $\lambda_{hs} \approx 4.1$. Inserting this in Eq.~\ref{eqVCSsxy} we obtain  $y_{\chi}  \approx \sqrt{2\lambda_s+4.3}$. Thus, because $\lambda_s \geq 0$, for stability reasons we need to have $y_{\chi} \gtrsim 2$.  This was the coupling used in Fig.~1 of Ref.~\cite{Frandsen:2013bfa} which constraints $m_s$ to be in the TeV range. Thus we conclude that the model  supports a 1st order EWPT and a viable DM candidate provided the VCs are not enforced. 


\subsubsection{Case (2): $\langle s(T_c) \rangle \neq$ 0, tree level scenario} 
\label{case2}

Now we shall consider the second case, where the tree level potential parameters satisfy the conditions 
\begin{eqnarray}
\label{s2c2}
\{ \mu^2, \lambda_h, -m^2_S, \lambda_{s}, \lambda_{hs} \} > 0 \quad \rm{with} \quad  0 < \lambda_h \lambda_{s} < \frac{\lambda^2_{hs}}{4} \, .
 \end{eqnarray}
The latter condition is needed to ensure that the potential does not provide a VEV to $S$   at zero temperature. 
There are also few other constraints which the parameters should satisfy in order for the model to support a strong 1st order EWPT. These are specified in connection with Eqs.~\ref{sxyc2P} - \ref{ms2} below.

Once the model parameters satisfy the conditions above, it is possible that at $T \neq 0$, the system has a configuration where the Higgs has a zero VEV, but the singlet has a non-zero VEV i.e.~the minimum of the potential is in the field space at point $(h,s)=(0,\langle s(T) \rangle)$, where $\langle s(T) \rangle \neq 0$. Following  \cite{Espinosa:2011ax}  in this case, for the EWPT study, we can use the zero temperature tree level potential supplemented with the leading finite-temperature terms.  
 
 More precisely, for a certain region of the parameter  space one can directly use the tree level potential to establish the presence of a strongly first order phase transition. This follows from the fact that here the potential can form two degenerate minima at positions $(h,s)=(0,\langle s(T_c) \rangle)$ and $(h,s)=(\langle h(T_c) \rangle,0)$ so that there is a  barrier, induced by the $\lambda_{hs} h^2 s^2$ interaction term, between the two minima.  Here the leading  high temperature expansion terms  are enough to trigger a first order phase transition between the finite temperature minima at $(0,\langle s(T_c) \rangle)$,  and $(\langle h(T_c) \rangle,0)$.  This potential structure for different temperatures is  elucidated in Fig.~\ref{fig:Veff}.

Intriguingly in this scenario there is a temperature range when a VEV occurs for the $S$ but not for $H$ leading to a phase in the evolution of the universe where the weak interactions are restored but the hyper-charge is still broken. There is, however, a new critical temperature above which also $S$ vanishes and all the symmetries are restored.

\begin{figure}
\begin{center}
\includegraphics[width=0.31\textwidth]{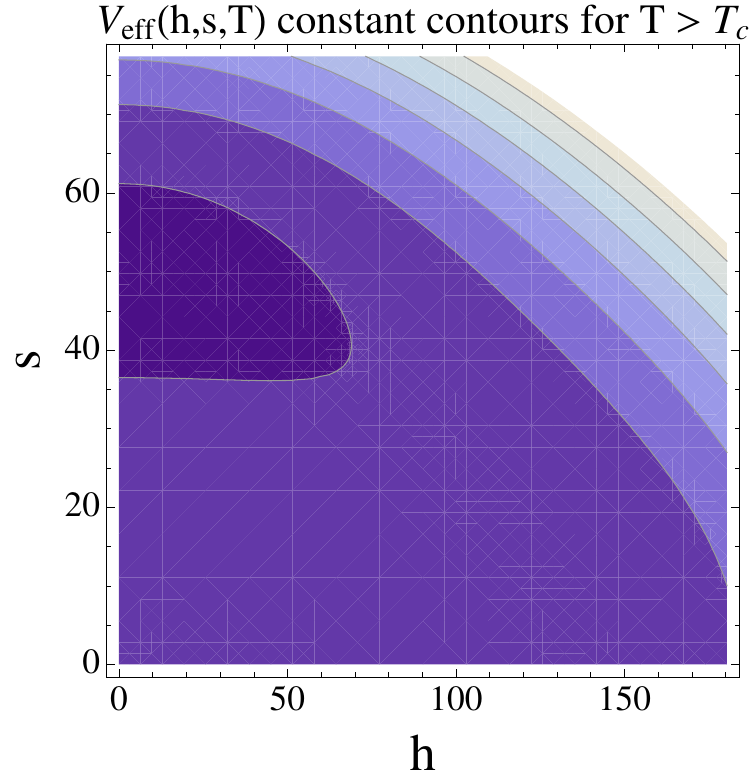} 
\includegraphics[width=0.31\textwidth]{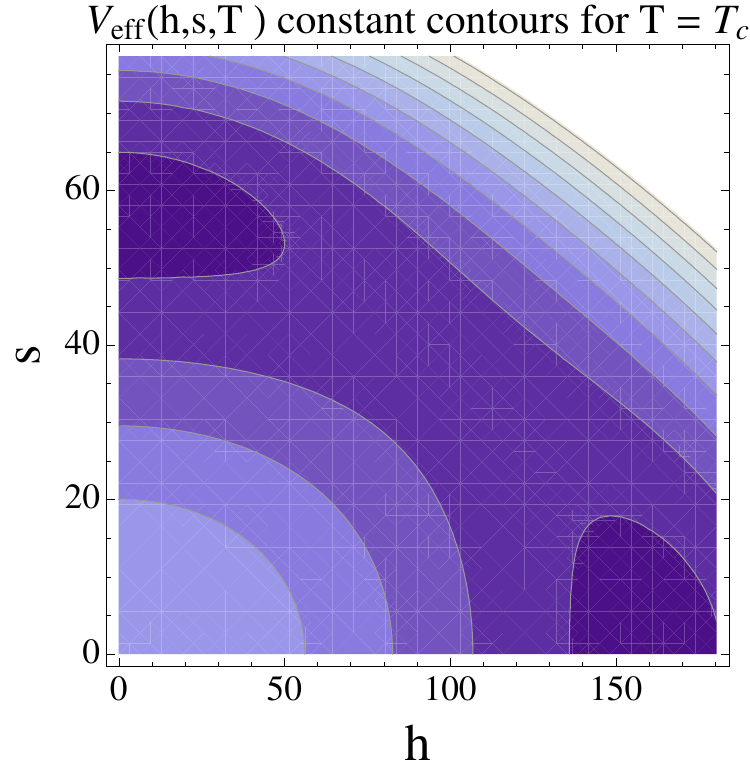} 
\includegraphics[width=0.32\textwidth]{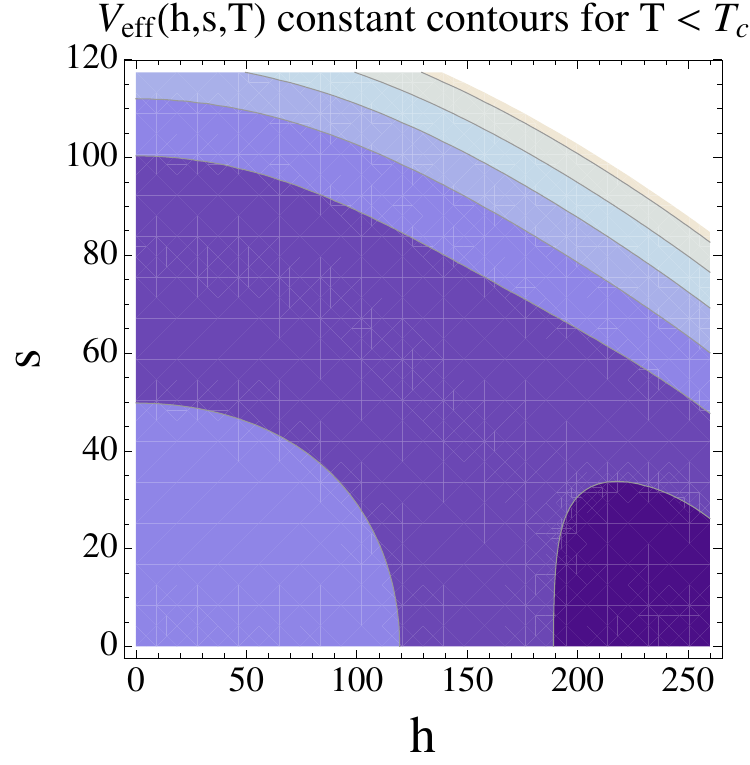} 
\caption{We show the finite temperature effective potential  contour plot for different temperatures: (Left) The "EW symmetry preserving" state with a minimum at $(h,s)=(0,\langle s(T) \rangle)$ at $T>T_c$; (Center) The  case with two degenerate minima $(h,s)= (h_c,s_c)$ at $T=T_c$;  (Right) The case where the EW symmetry is broken with a  minimum at $(h,s)=(\langle h(T) \rangle,0)$ for $T<T_c$. This shows the occurrence of a first order EWPT in the model.}
\label{fig:Veff}
\end{center}
\end{figure}
To establish the region of a strong first order EWPT we adopt the procedure described in \cite{Espinosa:2011ax}  and scan the parameters { using flat prior distributions} in the ranges
\begin{eqnarray}
\label{sxyc2P}
0.01 \leq \lambda_{hs} \leq 4 \pi,  \quad  0.01 \leq y_{\chi} \leq 4 \pi,  \quad  0<\lambda_{s,min} \leq \lambda_{s} \leq  \lambda_{s, max} < 4 \pi, 
 \end{eqnarray}
with the limits for $\lambda_s$ defined as
\begin{eqnarray}
\label{sxyc2P2} 
\lambda_{s,min} &\equiv& \frac{1}{2 a^2 \lambda_h} \left( c^2_h - 2 a \lambda_h  \delta c_s - c_h  \sqrt{c^2_h-4 a\lambda_h \delta c_s} \right), \quad 
\lambda_{s,max} \equiv \frac{\lambda_{hs}^2}{4 \lambda_h}, \\
c_s &\equiv& \Pi_s(T)/T^2, \quad \delta c_s \equiv c_s - a \lambda_s, \quad c_h \equiv  \Pi_h(T) /T^2, \\
 m^2_S &=& - \sqrt{\lambda_h \lambda_s} v^2_{EW} + \left( c_h \sqrt{\frac{\lambda_s}{\lambda_h}} -c_s  \right) T^2_c.
 \label{ms2}
 \end{eqnarray}
The parameter notation here follows closely the one used in \cite{Espinosa:2011ax} and we therefore also intruded the parameter $a$, which in our case has a value $a = 1/3$. 
The scalar thermal masses $\Pi_s(T)$ and $\Pi_h(T)$ are give in  Eq.~\ref{sxyTherMh} and Eq.~\ref{sxyTherMs} respectively of App.~\ref{Appsxy}. Further, in this setup, the   parameter $m_S$ can be written as a function of the critical temperature $T_c$ \cite{Espinosa:2011ax}. It is possible to randomly choose values for $T_c$ in the range $50$ GeV $\leq T_c \leq 250$ GeV and only accept the values providing $1 \leq h_c/T_c \leq 4$.  { Finally, $\lambda_h$ and $\mu$ are fixed at the SM values and the scan is done over $5 \times 10^5$ points.}  %
\begin{figure}[t]
\begin{center}
\includegraphics[width=0.30\textwidth]{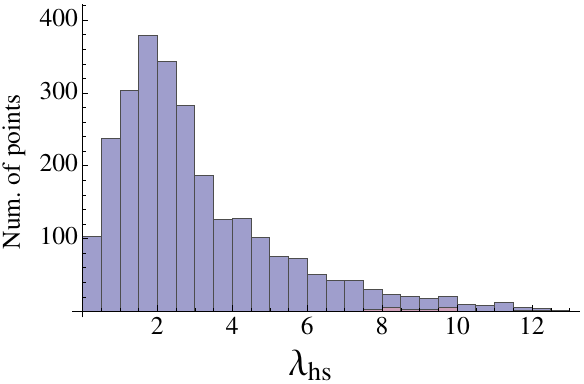} 
\includegraphics[width=0.30\textwidth]{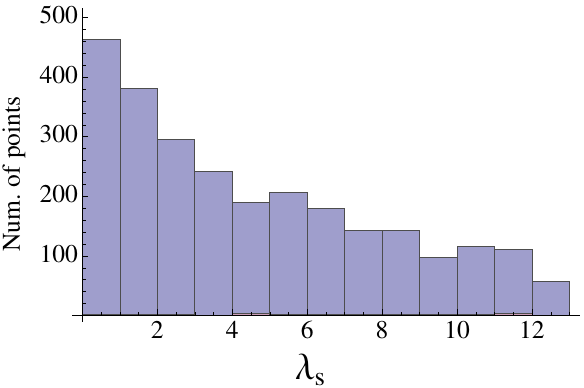} 
\includegraphics[width=0.30\textwidth]{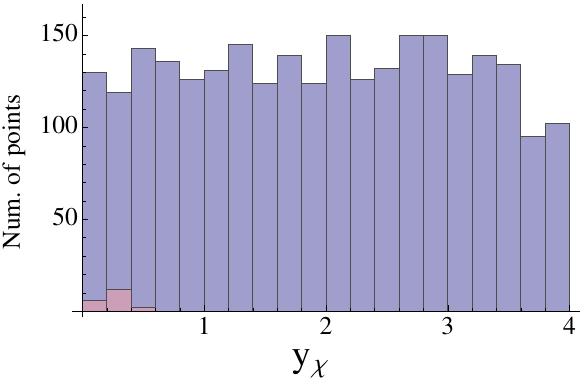} 
\includegraphics[width=0.30\textwidth]{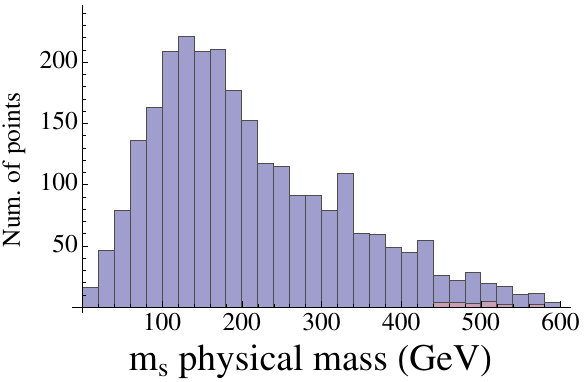} 
\includegraphics[width=0.30\textwidth]{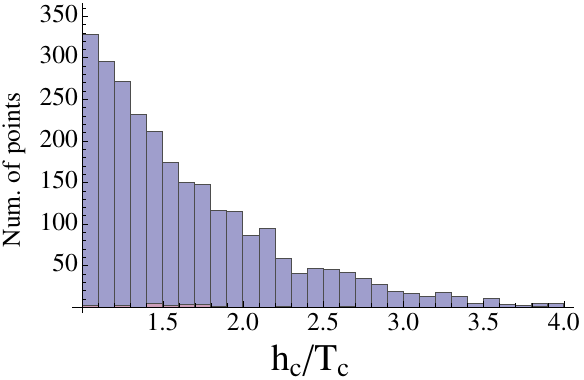} 
\includegraphics[width=0.30\textwidth]{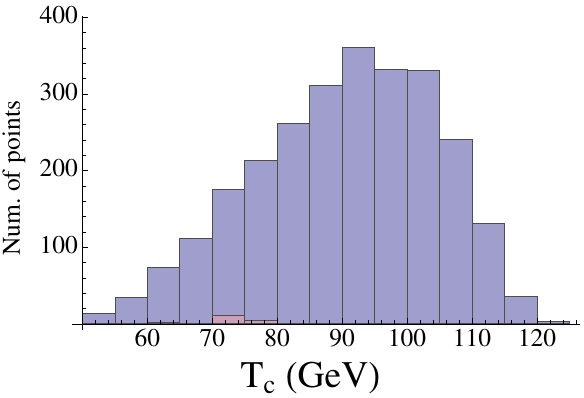} 
\includegraphics[width=0.30\textwidth]{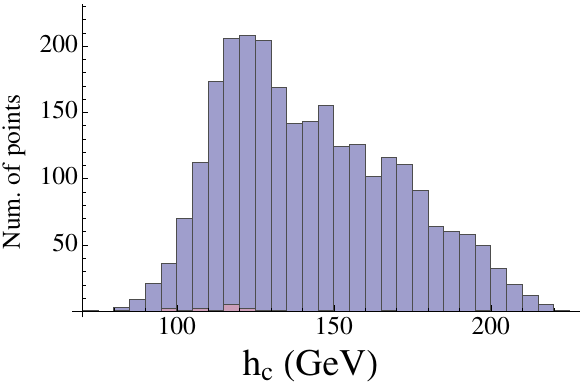} 
\includegraphics[width=0.30\textwidth]{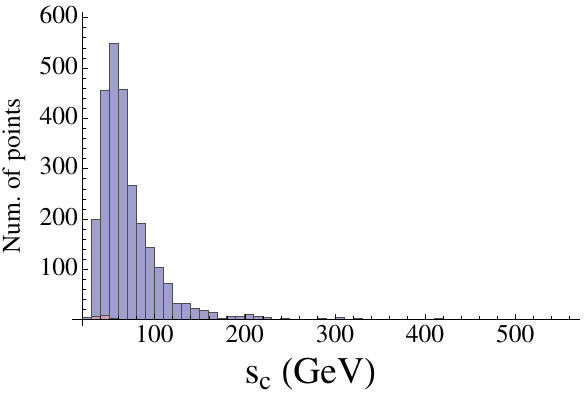} 
\includegraphics[width=0.30\textwidth]{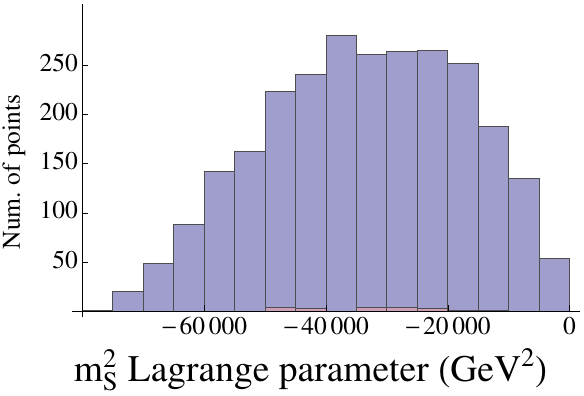} 
\caption{Distributions of  theSE$\chi$y model parameters in case (2) that produces a first order EWPT. The purple (blue) histograms show the distributions with(out) constraints $m_s >  450$ GeV and $y_{\chi} < 0.6$. The VCs have not been used here to constrain the parameters.}
\label{fig:sxy_c2}
\end{center}
\end{figure}
The distributions of the model parameters producing a first order EWPT is given in Fig.~\ref{fig:sxy_c2}. Purple (blue) histograms show the distributions with(out) the constraints $m_s >  450$ GeV and $y_{\chi} < 0.6$. These constraints were discussed in the end of 
 Sec.~\ref{sxyDM}. The results confirm the expectations that higher $S$ masses tend to reduce the order of the EWPT if we also ask the model to provide a phenomenologically viable DM candidate. Performing a full one loop analysis might slightly increases the purple region. 

\subsubsection{Veltman conditions}

Enforcing the VCs of Eq.~\ref{eqVCSsxy} and Eq.~\ref{eqVCHsxy}  requires, as in case (1),  $y_{\chi} \gtrsim 2$ that is above the LUX bound of  $y_{\chi} \lesssim 0.6$   allowing the scalar mass to be sufficiently light for the model to feature a first order EWPT.   
\\


 \subsection{Conformal SE$\chi$y Model} 
Our final model is the classically conformal limit of the SE$\chi$y model 
\begin{eqnarray}
\label{sxy_PNC}
   V_{0} & =&  \lambda_h (H^{\dagger}H)^2 +  \lambda_{hs}H^{\dagger}HS^{\dagger}S + \lambda_{s} (SS^{\dagger})^2+ (S \overline{E}{\chi}  y + \text{h.c.} ) \ .
\end{eqnarray}
We assume that only the Higgs field can get a non-vanishing VEV corresponding to the classically conformal limit of case (1)  in Sec.~\ref{sxyc1}. The CW and EWPT analyses are similar to the one performed for the real singlet scalar Higgs portal model scenarios and the quantities relevant for the EWPT are shown in App.~\ref{sexc1}.  

In  Fig.~\ref{fig:PNC_sxy_1} we show the $h_c /T_c$ contour plot as function of the SE$\chi$y Yukawa  coupling $y_{\chi} $ and the scalar S quartic self coupling $\lambda_{s}$. In the left plot, the portal coupling is fixed to $\lambda_{hs}\approx 6.85$ yielding the correct Higgs boson mass without VCs. The charged scalar $S$ mass is therefore also predicted to have a mass $m_{s} \approx 456$ GeV.   If we insist on  $\chi$ to be a thermal WIMP we need to have $m_E < m_{\chi} < m_s$, as argued in Sec.~\ref{sxyDM} \footnote{We allow here hard mass terms for $\chi$ and $E$. These mass terms could be generated radiatively via some yet new unknown dynamics.}. Because of the LHC constraints \cite{Dissauer:2012xa} we must also have  $m_E \gtrsim 393$ GeV. The DM mass is therefore constrained to be in a small mass window 393 GeV$\lesssim m_{\chi} \lesssim$ 456 GeV. This prediction is therefore testable at LHC run 2.  
\begin{figure}
\begin{center}
\includegraphics[width=0.43\textwidth]{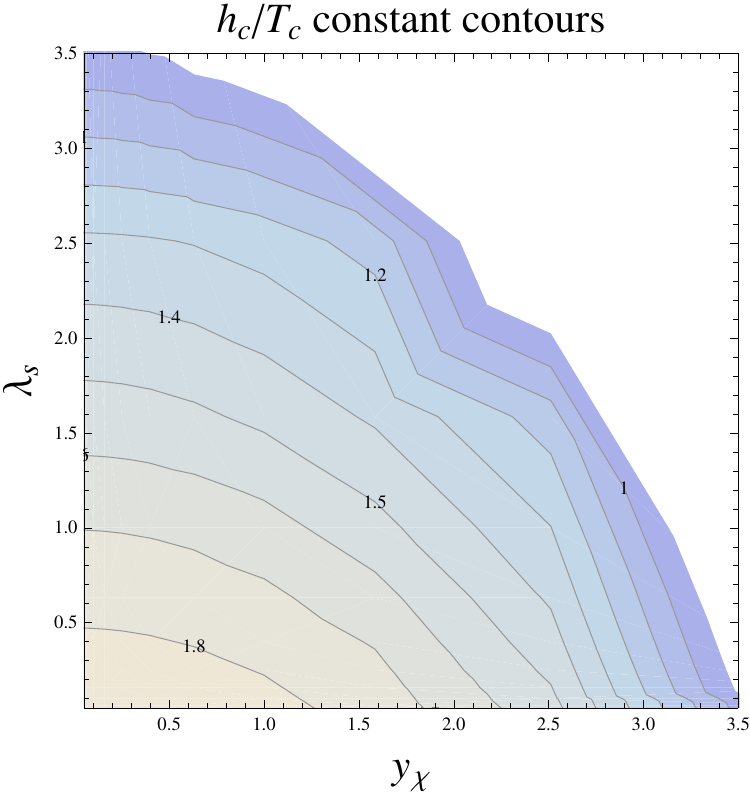} 
\includegraphics[width=0.42\textwidth]{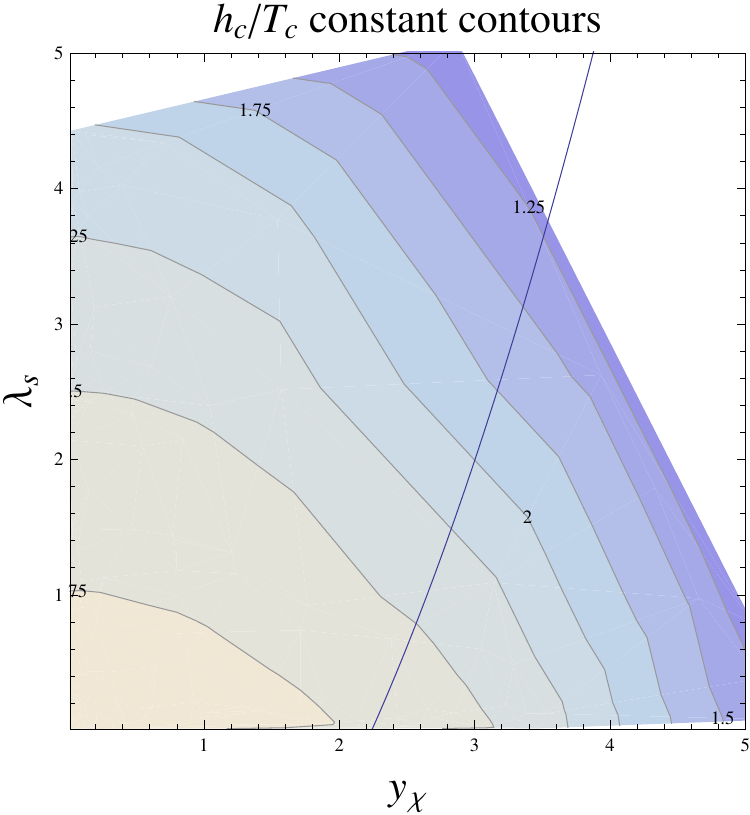} 
\caption{We shown the $h_c /T_c$ contour plots as function of the SE$\chi$y Yukawa  coupling $y_{\chi} $ and the scalar S quartic self coupling $\lambda_{s}$. 
In the left panel the portal coupling  is fixed at the value $\lambda_{hs}\approx 6.85$ that provides the correct Higgs boson mass without the VCs. On the right panel the portal coupling is fixed at the value $\lambda_{hs} \approx 4.84$, which satisfies the VC for the Higgs boson. The mass value of the Higgs is however not the physical one. On the upward turning blue curve the VC for the new scalar boson S mass is also satisfied.}
\label{fig:PNC_sxy_1}
\end{center}
\end{figure}

\subsubsection{Veltman conditions}

The Veltman conditions are again given by Eq.~\ref{eqVCSsxy} and Eq.~\ref{eqVCHsxy}. Using the flatness condition $ \lambda_h \approx 0$ at the EW scale, and the measured values for the rest of the known SM couplings in the Eq.~\ref{eqVCHsxy} we get that $\lambda_{hs} \approx 4.84$.
Using this value in Eq.~\ref{eqVCSsxy}  the VC for the new scalar relates $\lambda_s$ and $y_{\chi}$. This relation is plotted as an upward turning blue curve in Fig.~\ref{fig:PNC_sxy_1}. This figure is our result for the ratio $h_c /T_c$ as a function of  $y_{\chi} $ and  $\lambda_{s}$. 
The plot shows that the model can provide a strong first order phase transition when enforcing the VCs.  The prize to pay is a too small mass for the Higgs boson which now is  $m_h \approx 83$ GeV.


\section{Conclusions} 
\label{conc}

We analysed the EWPT of a large number of minimal extensions of the SM of particle interactions and their classically conformal limit. The models featured new scalars  and fermions and we uncovered the parameter space leading to a first-order phase transition with(out) the Veltman-Conditions. 

When relevant we discussed their dark (matter) aspects.  To accommodate simultaneously a first order EWPT and DM requires more involved extensions. 
{Our results for the classically conformal models are summarised in Table~\ref{tab:models}. We observe that classically conformal models can provide at most a subdominant DM component. Allowing for non-conformal terms, i.e.~adding extra operators with dimensionful couplings, one can saturate, in certain cases, the full cosmic DM density. In Table~\ref{tab:models2} we also summarise the results for few non-conformal model examples studied here. In the main text the reader will find further discussions of the results for the models summarised in the tables as well as for other minimal extensions not reported in the tables.}

A positive feature is that all these extensions must feature new particles lighter than a TeV making these models testable at the upcoming LHC run two experiments. 
  
\begin{table}[t]
\begin{center}
\begin{tabular}{|c|c|c|c|}
\hline
Conformal model & Higgs mass & $1^ {\rm st}$ order EWPT & DM   \\
\hline
{\rm SM} & - & + & - \\
{\rm SM + VC} & - & + & - \\\hline
{\rm RS; $\langle s \rangle =$ 0} & + & + & subdom. S  \cite{Cline:2013gha}\\
{\rm RS; $\langle s \rangle \neq$ 0} $^\#$& + & + & - \\
{\rm RS + VC; $\langle s \rangle \neq$ 0} $^\#$& - & + & -  \\\hline
{\rm RS + Wf; $\langle s \rangle =$ 0} & + & + & - \\
{\rm RS + Wf + VC; $\langle s \rangle =$ 0} & + & + & - \\
{\rm RS + Wf; $\langle s \rangle \neq$ 0} & + & + &subdom. Mf\\
{\rm RS + Wf + VC; $\langle s \rangle \neq$ 0} & + & + & - \\\hline
{\rm CS; $\langle s \rangle =$ 0} & + & + & subdom. S  \cite{Cline:2013gha}
\\\hline
{\rm CS + Df; $\langle s \rangle =$ 0} (SE$\chi$y)& + & + & - \\
{\rm CS + Df + VC; $\langle s \rangle =$ 0} (SE$\chi$y)& - & + & - \\\hline
{\rm CS + Mf; $\langle s \rangle =$ 0} \cite{Farzinnia:2014yqa} & + & + & subdom. PS  \cite{Farzinnia:2014yqa}
  \\\hline
\end{tabular}
\end{center}
\caption{ Summary of the different conformal models that can provide the observed Higgs boson mass, 1st order EWPT and DM. The Higgs portal models can feature Real Singlet (RS) with(out) a Weyl fermion(s) (Wf), or a Complex Scalar (CS) with(out) either a Dirac fermion(s) (Df) or Majorana fermion(s) (Mf).  We will also consider the cases with and without adding the Veltman Conditions (VC).
{ In these models the DM particle is identified as a scalar (S),  a pseudoscalar (PS) or a Majorana fermion.} The model, $\rm{RS + VC}; \langle s \rangle = 0$, is not included in the above list because the associated one-loop Higgs mass squared is negative. 
In the last column we also refer to the works where the results for the DM part were obtained. 
{ The results for the model in the last row are taken from \cite{Farzinnia:2014yqa}.}
Finally the symbol
$^\#$ means that the model suffers from the cosmic domain wall problem.}
\label{tab:models}
\end{table}  

\begin{table}[b!]
\begin{center}
\begin{tabular}{|c|c|c|c|}
\hline
Non-conformal Model & Higgs mass & $1^ {\rm st}$ order EWPT & DM   \\
\hline
{\rm SM} & + & - & - \\
{\rm SM + VC} $^\#$& + & - & - \\\hline
{\rm RS + Wf + VC; $\langle s \rangle =$ 0} (PNC)& + & + & + Mf\cite{Frandsen:2013bfa} \\
{\rm CS + Df; $\langle s \rangle =$ 0} (SE$\chi$y)& + & + & + Df \\
{\rm CS + Df + VC; $\langle s \rangle =$ 0} (SE$\chi$y)& - & + & + Df \\\hline
\end{tabular}
\end{center}
\caption{ Summary of the different non-conformal models that we have considered in this work and that can yield the observed Higgs boson mass, a 1st order EWPT and possibly a DM candidate { providing the full cosmic DM density}. We follow the notation introduced in Table~\ref{tab:models}, with the difference that here $^\#$ means that the model does not lead to the observed top quark mass.}
\label{tab:models2}
\end{table}  

\subsection*{Acknowledgments}
We thank Kimmo Kainulainen for discussions and comments on the manuscript. 
JV also thanks Matin Mojaza for discussions, and Tommi Alanne and Ville Vaskonen, for clarifying conversations related to their work.
The work has been partially supported by the Danish National Research Foundation under the grant DNRF:90. JV acknowledge the financial support from the Academy of Finland, projects 278722 and 267842.


\clearpage

\appendix
\centerline{\bf \Large Appendices A-F}
\vskip 1cm
We summarise in the following appendices the effective potential and other relevant quantities needed to determine the order of the EWPT for the various models investigated in the main text. 
{In the last appendix the relevant equations needed to determine the DM relic density for the PNC model are summarised. }

\section{Effective Potential}
\label{Effpot}

The full one-loop ring improved finite temperature effective potential, used in our perturbative study of the order of the EWPT, can be schematically written in the  form
\begin{eqnarray}
\label{potential}
V_{\rm Eff}(h,s,T) = V_0 (h,s) + V_1 (h,s) + V_T (h,s,T) + \delta V(h,s)  \,.
\end{eqnarray}
Here $V_0$ denotes the tree level potential, whereas $V_1$ and $ V_T$ are the one-loop zero and finite temperature parts of the potential respectively, and $\delta V$ is the counter term. We give details of these different terms below.
We will write the effective potential as a function of the background fields $(h,s)$, that are related to the Higgs doublet $H$ via $H = (0, h/ \sqrt{2})^T $, and to the new scalar field $S$ via $S = s/a$, where the normalization factor $a=1$ for real scalar and $a=\sqrt{2}$ for the complex scalar. 

We specify the tree level potential $V_0$ for each model separately at the beginning of each section in the main text.  To be precise, we implicitly take that $V_0 (h,s) \equiv V_{\rm SM} (h) + V_0 (h,s)$, where $V_{\rm SM}$ is the SM part of the potential, except for the pure scalar part of the potential, which is given in $V_0$. Also the non-SM contributions to the potential are defined in $V_0$.

The one-loop zero temperature contribution to the effective potential is the Coleman-Weinberg one
\begin{eqnarray}
\label{V1}
V_1 (h,s) = \sum_{i} \pm \, n_i  \frac{m^4_i(h,s)}{64 \pi^2 } \left( \log \left[ \frac{m^2_i(h,s)}{Q^2} \right] -C_i  \right )  \,,
\end{eqnarray}
where the sum runs over the bosons (+) and fermions ($-$) and $n_i$ counts the internal degrees of freedom of each species $i$. 
The background field dependent tree level masses for each field are denoted by $m_i(h,s)$, $Q$ is the renormalization scale, $C_i$ is a constant equal to 5/6 for gauge bosons and 3/2 for scalar bosons and fermions. 
The relevant degrees of freedom and specific mass terms $m_i(h,s)$ are given for each model separately in App.~\ref{SMmass}, \ref{AppHS} and \ref{Appsxy}.
 
 The counter term part of the potential can be written in the form 
\begin{eqnarray}
\label{Vcount}
\delta V (h,s) = \frac{\delta \mu^2}{2} h^2 + \frac{\delta \lambda_h}{4} h^4 +
 \frac{\delta \lambda_{hs} }{4} h^2 s^2 +\frac{\delta m_s^2}{2} s^2+ \frac{\delta \lambda_{s}}{4}s^4 + \delta V_0 \, ,
\end{eqnarray}
where the counter-term parameters $\delta \mu^2, \delta \lambda_h,\delta \lambda_{hs},\delta m_s^2,\delta \lambda_{s}$, and $ \delta V_0$ are fixed by the renormalization conditions. It is more practical to define these conditions for the classically conformal and non-conformal models separately. 

Among the non-conformal SM extensions we have studied in detail the "SE$\chi$y" model, see Sec.~\ref{sxymodel}.  For case (1) of this model presented in  Sec.~\ref{sxyc1}, the renormalization conditions are similar to that of the SM i.e.~only $\delta \mu^2$ and $\delta \lambda_h$ are relevant and need to be fixed by requiring that the minimum of the effective potential and the Higgs boson mass value are held at their tree level values, that is
\begin{displaymath}
\frac{\partial}{\partial h} \left( V_1 (h,s)  + \delta V (h,s) \right)\Big|_{\begin{subarray}{l} h=v_{EW} \\ s=0 \end{subarray}} =0 \quad \rm{and}  \quad 
\frac{\partial^2}{\partial h^2} \left( V_1 (h,s) + \delta V (h,s) \right) \Big|_{\begin{subarray}{l} h=v_{EW} \\ s=0 \end{subarray}} =0\, .
\end{displaymath}

In case (2), studied in Sec.~\ref{case2}, because we are using only the tree level potential in the EWPT study (with leading finite T contributions), we do not need to determine the counter-terms. The renormalization conditions for the conformal models are specified in Sec.~\ref{flat}.

The one-loop finite temperature part reads
\begin{eqnarray}
\label{VT}
V_T (h,s,T) & = & \sum_{i} \pm \, n_i \frac{T^4}{2 \pi^2 } \int_{0}^{\infty} {\rm{d}} y y^2 \log{ \left[ 1 \mp \exp{ -\sqrt{y^2 + m^2_i(h,s)/T^2} } \right] } \\
                     & + & \delta_{i,B}  \frac{T}{12 \pi}\left[ m^3_i(h,s) -  \mathrm{m}^3_i(h,s,T) \right ] ,
\end{eqnarray}
where again (+) refers to bosons and ($-$)  to  fermions. The last term including the Kronecker delta function takes into account the usual ring improvement (daisy diagram resummation) for bosonic (B) degrees of freedom, where $\mathrm{m}_i(h,s,T)$ stands for the thermally corrected boson mass. To speed up the calculations  we used an approximation for the integral in Eq.~(\ref{VT}) \cite{Li:2014wia}.  This approximation consists in smoothly joining the high and low temperature expansion of the integral and is summarized in App.~\ref{AppA}. 

 \subsection{Classically conformal case}
\label{flat}

In the classically conformal cases the EW symmetry breaking needs to be triggered by loop effects i.e.~via dimensional transmutation $\rm \grave{a}~la$ Coleman-Weinberg (CW) \cite{Coleman:1973jx}. 
For this to happen the tree level scalar potential needs to have a flat direction.
The flat direction conditions, for generic scalar potentials has been provided in\cite{Gildener:1976ih}. Particularly, we follow the flat direction and CW analysis done in \cite{Antipin:2013exa} for the models studied here, which relies on   \cite{Gildener:1976ih,Coleman:1973jx}. We shall briefly summarise the core points here and end with the renormalization conditions for the classically conformal case.

Let us start with classically conformal tree level scalar potential
\begin{eqnarray}
V =   \frac{\lambda_{ijkl}}{24}\phi_i \phi_j \phi_k  \phi_l + c.t.\ ,
\end{eqnarray}
where the $\lambda_{ijkl}$ are renormalized couplings and $c.t.$ refers to counter terms. In principle the model can (and will) feature also fermionic and vectorial contributions but for our purposes to find the flat directions in the pure scalar potential it is enough to consider the previous form of the potential and take that $\lambda_{ijkl} \sim g^2 \sim y^2 \ll 1$, where $g$ and $y$ refers to possible gauge and Yukawa couplings included in the model. For the CW mechanism to be operative the model scalar quartic couplings need to be the such that there is a flat direction at tree level.
Particularly the couplings $\lambda_{ijkl}$ should be renormalized such that at the renormalization scale $Q$ the potential has indeed a flat direction. To this end, let $u_i$ be a unit vector in the scalar field space, then to find the flat direction we need to solve 
\begin{eqnarray}
{\rm{min}} \left( \lambda_{ijkl} u_i u_j u_k u_l \right)|_{u_i u_i=1} =   0 \,.
\end{eqnarray}
If a solution $u_i = n_i$ can be found, then $\phi_i = n_i \phi$ defines a flat direction in tree level potential. Thus the couplings should be renormalized such that
\begin{eqnarray}
\label{remin}
\lambda_{ijkl}(Q) n_i n_j n_k n_l  \approx   0 \,.
\end{eqnarray}

Now we shall write the potential using a classical background field such that  $\phi_{cl} = n_i \phi_i$. For our analysis it is convenient to use polar coordinates in the scalar field space. The
tree level potential acquires the form 
\begin{eqnarray}
V(h,s)_0 = V(r, \omega)_0 = r^4 f(\omega),
\end{eqnarray}
where the function $f(\omega)$ can be read from the original potential $V(h,s)_0$ once substituting $h =r \cos \omega$ and $s =r \sin \omega$.
As usual, the minimum can be found by studying the first and second derivatives of the potential. Specifically the model parameters satisfying the condition $f'(\omega)=0$ define the direction $\langle \omega \rangle$ of the non-trivial minimum as discussed in Sec.~\ref{SMS_c2}.
Now the unit vector defining the flat direction is $n = (\cos \langle \omega \rangle, \sin \langle \omega \rangle)$.  And the flat direction problem in our case takes the form
\begin{eqnarray}
{\rm{min}} \left( \lambda_{ijkl} u_i u_j u_k u_l \right)|_{u_i u_i=1} = {\rm{min}} f(\omega) |_{u_i u_i=1}  =  0 \,.
\end{eqnarray}
As we are using the polar coordinates the background field is now $\phi_{cl} = r$, and the one loop potential $V_1 (h,s)$  can be written in the form
\begin{eqnarray}
V_1 (\phi_{cl}, \omega)   &=&  \sum_{i} \pm \, n_i  \frac{ \phi^4_{cl} W^4_i(\omega)}{64 \pi^2 } \left( \log \left[ \frac{\phi^2_{cl} W^2_i(\omega)}{Q^2} \right] -C_i  \right )  \\
&=& A \phi^4_{cl} + B  \phi^4_{cl} \log \left[ \frac{\phi^2_{cl}}{Q^2} \right] \,,
\label{Vcw}
\end{eqnarray}
where 
\begin{eqnarray}
A \equiv \sum_{i} \pm \, n_i  \frac{W^4_i(\omega)} {64 \pi^2 }  \left( \log [ W^2_i(\omega)] -C_i  \right )  \quad \rm{and} \quad  B \equiv \sum_{i} \pm \, n_i  \frac{W^4_i(\omega)} {64 \pi^2 } 
\end{eqnarray}
and we have decomposed all the mass terms in the form $m^2_i(\phi_{cl}, \omega) = W^2_i(\omega) \phi^2_{cl}$ using the fact that in the classically conformal case all the mass terms are directly proportional to  $\phi_{cl}$. The quantity $W^2_i(\omega) $ thus depends only on the model couplings since in the end also $\omega$ will depend only on the scalar quartic couplings.

The full zero temperature potential now reads $V(\phi_{cl}, \omega) = V_0(\phi_{cl}, \omega) + V_1(\phi_{cl}, \omega) + \delta V (\phi_{cl}, \omega)$. And finally, the renormalization conditions for quartic couplings, fixing the counter terms, is defined via relation
\begin{eqnarray}
\frac{\partial^4 V}{\partial \phi^4_{cl} } \bigg{|}_{m^2_i(\phi_{cl},\omega)=Q^2 e^{(-25/6+C_i)}}= \lambda_{ijkl}(Q) n_i n_j n_k n_l  =   0 \,,
\end{eqnarray}
with the renormalization scale chosen to be $Q \sim v_{EW}$. The factor $e^{-25/6}$ is extracted for convenience and follows the renormalization choice made in \cite{Gildener:1976ih}.
For completeness, let us also give the relation for the scalar (Higgs) mass in this case. 
The non-trivial extremum of the potential Eq.~\ref{Vcw} is at  
\begin{eqnarray}
 \log \frac{\langle \phi^2_{cl} \rangle}{Q^2 } = -\frac{1}{2} -\frac{A}{B} \,,
\end{eqnarray}
and if this corresponds to a ground state, the scalar fluctuation along the flat direction has a positive mass squared 
\begin{eqnarray}
 m^2_{CW} = 8 B  \langle \phi^2_{cl} \rangle \,,
\end{eqnarray}
which defines the CW scalar boson mass.

\section{Thermal integral} 
\label{AppA}

Here we give the approximation we used for the thermal integral, Eq.~\ref{VT} that  smoothly joins the high and low temperature expansions of the thermal integral. This approximation has been used previously in \cite{Li:2014wia}.
The thermal integral in Eq.~\ref{VT} is of the form 
\begin{eqnarray}
\label{int}
I_{B,F} (a) & = & \pm \, \int_{0}^{\infty} {\rm{d}} y y^2 \log{ \left[ 1 \mp e^{ -\sqrt{y^2 + a} } \right] } \, ,
\end{eqnarray}
where the upper sings refer to bosons and the lower to fermions and $a \equiv m^2_i(h,s)/T^2$.
The high temperature expansions $(m_i/T \ll 1)$ of the integral, Eq.~\ref{int}, for the bosons and fermions can be written as \cite{Dolan:1973qd}
\begin{eqnarray}
I^H_{B} (a) & = & -\frac{\pi^4}{45}+ \frac{\pi^2}{12}a -\frac{\pi}{6}a^{\frac{3}{2}}-\frac{a^2}{32}(\log(a) -c_B)  \, ,\\
I^H_{F} (a) & = &  -\frac{7 \pi^4}{360}+ \frac{\pi^2}{24}a+\frac{a^2}{32}(\log(a) -c_F) \, ,
\end{eqnarray}
where $c_B = 3/2 - 2 \gamma_E + 2 \log (4\pi)$ and $c_F =3/2 - 2 \gamma_E + 2 \log (\pi) $ and $\gamma_E \approx 0.5772$.
The low temperature expansion $(m_i/T \gg 1)$ of the integral is the same both for  bosons and fermions and reads \cite{Anderson:1991zb}
\begin{eqnarray}
I^L_{B,F} (a) & = & - \sqrt{\frac{\pi}{2}}  a^{3/4} e^{-a^{\frac{1}{2}}}\left(1+ \frac{15}{8}a^{-\frac{1}{2}} +\frac{105}{128}a^{-1} \right) .
\end{eqnarray}
The approximated expression $I^A_{B,F} $ for the thermal integral, $I_{B,F} $ reads
\begin{eqnarray}
I^A_{B,F} (a) & = & t_{B,F}(a) I^H_{B,F}(a)+(1- t_{B,F}(a)) I^L_{B,F} (a) \, \approx  \, I_{B,F}(a) \, ,
\end{eqnarray}
where the fitting functions for bosons and fermions, that join the high and low temperature expressions, are $t_{B}(a)=e^{-(a/6.3)^4}$ and $t_{B}(a)=e^
{-(a/3.25)^4}$ respectively  \cite{Li:2014wia}. The relative error in this approximation at worst is less than 5$\%$ in the range $0 \le a \le 20$. 
Finally, the approximated finite temperature part of the effective potential is
\begin{eqnarray}
V^A_T (h,s,T) & = & \frac{T^4}{2 \pi^2} \left[ \sum^B_{i} n_i I^A_{B} (m^2_i(h,s)/T^2) + \sum^F_{i} n_i I^A_{F} (m^2_i(h,s)/T^2)\right ] \\
                         & + & \sum^B_{i} n_i \frac{T}{12 \pi}\left[ m^3_i(h,s) -  \mathrm{m}^3_i(h,s,T) \right ]   \, ,
\end{eqnarray}
where $n_i$ counts the internal degrees of freedom and in the second line we have added the thermal ring improvement for the bosonic degrees of freedom. 
 
\section{Mass Terms for the SM} 
\label{SMmass}

Here we give the relevant mass terms used in the one-loop effective potential, Eq.~\ref{potential}, when studying the EWPT in the SM. 
These terms are also relevant for the other models considered in this work. 
The background field dependent masses, discussed in connection with Eq.~\ref{potential}, are of the form $m_i(h,s) = m_i(h)$ and $m_i(h,s,T) = m_i(h,T)$, because the only background field in this case is the Higgs mean field $h$.
Now, the well known $h$ dependent tree level mass terms in the SM are 
\begin{eqnarray}
\label{SMTreeM}
&& m^2_{h_0}(h) = 3 \lambda_h h^2- \mu^2, \quad
   m^2_{G}(h) = \lambda_h h^2- \mu^2, \, \\ \label{SMTreeMT}
&& m_{W,T}(h) =  m_{W,L}(h) =\frac{g}{2} h, \quad
   m_{Z,T}(h) = m_{Z,L}(h) =\frac{\sqrt{g^2+g'^2}}{2} h, \\
&& m_{t}(h) = \frac{y_t}{\sqrt{2}} h, 
  \end{eqnarray}
where $m_{h_0}$ and  $m_{G}$ are the Higgs boson and the Goldstone boson masses respectively. We have divided the gauge boson masses $m_{W,i}$ and $m_{Z,i}$ into transverse $(i=T)$ and longitudinal $(i=L)$ components for later use. Finally, $m_{t}$ is the top quark mass. As mentioned before, we only take into account the top quark contribution from the SM fermions. For completeness, we write down also the Goldstone boson terms, even though we do not consider them in the analysis, as discussed in the end of Sec.~\ref{intro}. 
The internal degrees of freedom for different species are
\begin{eqnarray}
\label{SMDF}
&& n_{h_0}= 1,\quad
    n_{G} = 3  \, ({\rm three \, Goldstones}), \, \\
&&  n_{W,T} =4 \, ({\rm polarization  \, \times \, particle/antip. }), \quad
   n_{W,L} =2, \quad
   n_{Z,T} = 2, \quad
   n_{Z,L} = 1,\quad \, \\
&&   n_{t} = 12  \,   ({\rm spin \, \times \, particle/antip. \, \times \, color}), 
\end{eqnarray}
where we have used the same subscript notation as for the masses above.
The scalar boson thermal masses are
\begin{eqnarray}
\label{SMTherM}
 \mathrm{m}^2_{h_0}(h,T) &=& m^2_{h_0}(h) + \Pi^{SM}_h(T), \\
 \mathrm{m}^2_{G}(h,T) &=& m^2_{G}(h) + \Pi^{SM}_h(T),
 \end{eqnarray}
where the thermal self energy (at momentum $p=0$ limit) for the Goldstone and for Higgs boson is the same  
\begin{eqnarray}
\label{SMTherMh}
 \Pi^{SM}_{h}(T) &=& \frac{1}{16} \left[  3 g^2 + g'^2+ 8 \lambda_h + 4 y_t^2 \right] T^2  .
 \end{eqnarray}
Finally, the longitudinal gauge boson thermal masses follows by diagonalizing the mass matrix
\begin{eqnarray}
\label{SMGM}
 \mathrm{M}^2_L(h,T) =  \left( \begin{array}{cccc} g^2 & 0      & 0    &   0  \\
				                                       0 & g^2 & 0    &   0 \\
				                                       0 &   0   & g^2&   - g g' \\
				                                       0 &   0   & - g g'&   g'^2  \end{array} \right) \frac{h^2}{4} + 	
                            \left( \begin{array}{cccc} g^2 & 0      & 0    &   0  \\
				                                       0 & g^2 & 0    &   0 \\
				                                       0 &   0   & g^2&   0\\
				                                       0 &   0   &    0 &   g'^2  \end{array} \right) \frac{11}{6}T^2   , 
 \end{eqnarray}
where the two upper rows refers to W-boson contributions, and the other two  to the Z-boson-photon mixing contributions. 
The gauge boson transverse components do not get thermal corrections at the one-loop level, and thus the transverse thermal masses are equal to the tree level masses i.e.~$m^2_{i,T}(h,T)=m^2_{i,T}(h)$ for $i=W,Z,\gamma$ with $m^2_{i,T}(h)$:s given in Eq.~(\ref{SMTreeMT}) except for photon, which is $m_{\gamma,T}(h)=0$.


\section{Mass Terms in the Extension with A Real Singlet Scalar} 
\label{AppHS}

\subsection{Case (1) and non-conformal scenario}
\label{AppHSc1}

In this case the singlet scalar does not acquire a non-zero VEV and the EWPT analysis is similar to the SM case.  {The difference with respect to the SM case resides in the corrections to the SM scalar boson mass terms and the addition of the singlet scalar effective potential. The latter is effectively obtained, i.e.~the part contributing to the EWPT, by adding the singlet in the sums in Eq.~\ref{V1} and Eq.~\ref{VT} with 
a tree level mass term $m^2_{s_0}(h) = m^2_S + \lambda_{hs} h^2 $, and one internal degree of freedom, i.e. $n_{s} = 1$.}
The scalar boson thermal masses are now
\begin{eqnarray}
\label{SMSTherM}
 \mathrm{m}^2_{h_0}(h,T) &=& m^2_{h_0}(h) + \Pi_h(T),\\
\mathrm{m}^2_{s_0}(h,T) &=&  m^2_{s_0}(h)  + \Pi_s(T), \\
 \mathrm{m}^2_{G}(h,T) &=& m^2_{G}(h) + \Pi_h(T), \, 
 \end{eqnarray}
where the thermal self energies (at momentum $p=0$ limit) are  
\begin{eqnarray}
\label{SMSTherMh}
 \Pi_h(T) &=& \Pi^{SM}_{h}(T) + \frac{\lambda_{hs}}{12} T^2 , \\
  \label{SMSTherMs}
 \Pi_s(T) &=& \left[  \frac{ \lambda_{hs}}{3}+ \frac{\lambda_{s}}{4} \right]  T^2 ,
 \end{eqnarray}
with $\Pi^{SM}_{h}(T)$  given in Eq.~\ref{SMTherMh}. Notice also that there is no mixing between $h$ and $s$ in this case. Thus, in the end, the physical Higgs field is directly related to $h$.

In the classically conformal case, i.e.~case (1), the parameters are as above except that $\mu^2$ and $m^2_S$ are set to zero. 

\subsection{Case (2)}
\label{AppHSc2}

Here the singlet scalar acquires a non-zero VEV, the mass terms given above are therefore partially modified. The tree level scalar mass terms are obtained by diagonalizing the mass matrix  
\begin{eqnarray}
\label{SMrSc2}
M^2_{h_0,s_0}(h,s) =  \left( \begin{array}{cc}  \lambda_h 3 h^2 +  \lambda_{hs} s^2 & 2 \lambda_{hs} h s \\
				                                    2 \lambda_{hs} h s  &  \lambda_s 3 s^2 +  \lambda_{hs}  h^2 \end{array} \right) ,
\end{eqnarray}
and the scalar thermal masses by diagonalizing the matrix
\begin{eqnarray}
\label{SMrSc2term}
 \mathrm{M}^2_{h_0,s_0}(h,s,T) = M^2_{h_0,s_0}(h,s)  + \left( \begin{array}{cc}  \Pi_{h}(T)  & 0  \\
0  & \Pi_{s}(T) \end{array} \right)  .
\end{eqnarray}
where the thermal self energies are given in Eq.~\ref{SMSTherMh} and Eq.~\ref{SMSTherMs}.
All the other mass terms are like in the SM, with $\mu^2$ and $m^2_S$ set to zero.

As discussed in the main text and in App.~\ref{flat}, for the CW and EWPT analysis it is convenient to use polar coordinates $(r,\omega)$ in the scalar field space.
We get the mass parameters in these coordinates by substituting $h \rightarrow r \cos  \omega $, and, $s \rightarrow r \sin \omega $.


\subsection{Mass terms in PNC Model}

\subsubsection{Case (1)}
\label{AppPNCc1}

In this case the singlet scalar does not acquire a VEV and therefore the relevant terms are like in case (1) considered in Sec.~\ref{AppHSc1}, except for the extra contribution following from the singlet Weyl fermion. Specifically, the only modification is in the singlet scalar thermal self energy (at momentum $p=0$ limit), which now reads
\begin{eqnarray}
  \label{SMSschi}
 \Pi_s(T) &=& \left[  \frac{ \lambda_{hs}}{3}+ \frac{\lambda_{s}}{4} + \frac{1}{3}y_{\chi}^2 \right]  T^2 .
 \end{eqnarray}

\subsubsection{Case (2)}

Here also the singlet scalar receives a VEV, in this case the mass parameters are like in case (2) considered in Sec.~\ref{AppHSc2} above, except for the (Weyl) Majorana fermion $\chi$ that receives a tree level mass $m_{\chi}(s) = 2 y_{\chi} s $, and should be taken into account in effective potential, with $n_{\chi} = 2$. Of course we need to take into account also of the singlet scalar thermal self energy $ \Pi_s(T) $ given in Eq.~\ref{SMSschi}.


\section{Mass Terms in SE$\chi$y Model} 
\label{Appsxy}


\subsection{Case (1) and conformal scenario}
\label{sexc1}

In this case the scalar $S$ does not acquire a VEV at any temperature. 
Thus the EWPT follows closely the SM and the real singlet Higgs portal scenarios, in cases (1), considered above.  
As in the case in Sec.~\ref{AppHSc1}, here the only background field is the $h$ field, and the background field dependent masses are of the form $m_i(h,s) = m_i(h)$ and $m_i(h,s,T) = m_i(h,T)$. 
All terms related to the SM particles are just like in the Sec.~\ref{SMmass} above, except for the mass terms which we shall refine here. 
Thus just like in the case of SM + real singlet, addition to the SM contributions to the 1-loop effective potential, there is now extra piece coming from the $S$ scalar, with tree level mass term $m^2_{s}(h) = m^2_S + \frac{\lambda_{hs}}{2} h^2 $, with internal degrees of freedom $n_{s} = 2$ for complex scalar. 
The scalar boson thermal masses are
\begin{eqnarray}
\label{sxyTherM}
 \mathrm{m}^2_{h}(h,T) &=& m^2_{h}(h) + \Pi_h(T),  \\
\mathrm{m}^2_{s}(h,T) &=&  m^2_{s}(h)  + \Pi_s(T), \\
 \mathrm{m}^2_{G}(h,T) &=& m^2_{G}(h) + \Pi_h(T), 
 \end{eqnarray}
where the thermal self energies (at momentum $p=0$ limit) are  
\begin{eqnarray}
\label{sxyTherMh}
 \Pi_h(T) &=& \Pi^{SM}_{h}(T) + \frac{\lambda_{hs}}{12} T^2, \\
  \label{sxyTherMs}
 \Pi_s(T) &=& \left[  \frac{ \lambda_{hs}}{6}+ \frac{\lambda_{s}}{3}+ \frac{1}{4} g'^2+ \frac{1}{12}y^2 \right]  T^2, 
 \end{eqnarray}
with $\Pi^{SM}_{h}(T)$ like in the SM in Eq.~\ref{SMTherMh}.
Notice that there is no mixing between $h$ and $s$ in this case. Thus the physical Higgs field is associated directly with $h$. 
Finally, the longitudinal gauge boson thermal masses we get by diagonalizing the mass matrix
\begin{eqnarray}
\label{sxyGM}
 \mathrm{M}^2_L(h,T) =  \left( \begin{array}{cccc} g^2 & 0      & 0    &   0  \\
				                                       0 & g^2 & 0    &   0 \\
				                                       0 &   0   & g^2&   - g g' \\
				                                       0 &   0   & - g g'&   g'^2  \end{array} \right) \frac{h^2}{4} + 	
                            \left( \begin{array}{cccc} g^2 & 0      & 0    &   0  \\
				                                       0 & g^2 & 0    &   0 \\
				                                       0 &   0   & g^2&   0\\
				                                       0 &   0   &    0 &   g'^2 (1+\frac{13}{24}\frac{6}{11} )  \end{array} \right) \frac{11}{6}T^2  , 
 \end{eqnarray}
where the two upper rows refers to W-boson contributions, where as the two lower to the Z-boson-photon mixing contributions. The new contributions to the B-field thermal mass follows from the new charged fields: from the scalar $S$ and from the vector like electron $E$.
The transverse thermal masses are as in the SM case i.e.~$m^2_{i,T}(h,T)=m^2_{i,T}(h)$ for $i=W,Z,\gamma$ with $m^2_{i,T}(h)$:s given in Eq.~(\ref{SMTreeMT}) except for photon which is $m_{\gamma,T}(h)=0$.

In the classically conformal scenario, the mass terms are as above, expect that the dimension full scalar Lagrange mass parameters, $\mu^2$ and $m^2_S$, are set to zero everywhere.


\subsection{Case (2)} 
\label{Appsxyc2}

{ Following the approach of \cite{Espinosa:2011ax}, for the {\it tree level} strong 1st order EWPT calculations, in addition to the tree level potential, we need only the one-loop scalar thermal masses. These are given already in Eq.~\ref{sxyTherMh} and Eq.~\ref{sxyTherMs} above. Further, the tree level model parameters should satisfy the relations given in Eq.~\ref{s2c2} related to Eqs.~\ref{sxyc2P} - \ref{ms2} in Sec.~\ref{case2}.}

For completeness, we also give the mass terms in a form that can be used for a full one loop EWPT calculation. 

Because we have two background fields, $h$ and $s$, the mass terms depend on both, i.e.~$m_i(h,s) $ and $m_i(h,s,T)$. 
We have also mixing between the $h$ and $s$ fields here and thus we write the mass terms in matrix form. We deduce the tree level scalar masses by diagonalizing the scalar mass matrix 
\begin{eqnarray}
\label{sxySMc2}
M^2_{h,s}(h,s) =  \left( \begin{array}{cc}  \lambda_h 3h^2- \mu^2 + \frac{\lambda_{hs}}{2} s^2 & \lambda_{hs} h s  \\
				                                     \lambda_{hs} h s  &  \lambda_{s} 3s^2-m^2_S + \frac{\lambda_{hs}}{2} h^2 \end{array} \right).
\end{eqnarray}
The Goldstone boson masses are
\begin{eqnarray}
\label{sxyTreeM}
M^2_{G,\sigma}(h,s) =  \left( \begin{array}{cc}  \lambda_h h^2- \mu^2 +\frac{\lambda_{hs}}{2} s^2 & 0  \\
				                                     0  &  \lambda_{s} s^2-m^2_S + \frac{\lambda_{hs}}{2} h^2 \end{array} \right) ,
\end{eqnarray}
where the first row, associated with $G$, stands for all the three SM Goldstone boson masses, whereas the lower row, refers to $\sigma$, that is the mass of the Goldstone boson associated with the broken U(1) symmetry. This occurs because the electrically charged scalar $S$ acquires a background field dependent VEV.  The Thermal masses of the scalars are 
\begin{eqnarray}
\label{sxySTMc2}
 \mathrm{M}^2_{h,s}(h,s,T) = M^2_{h,s}(h,s) + \left( \begin{array}{cc}  \Pi_{h}(T)  & 0  \\
0  & \Pi_{s}(T) \end{array} \right)  .
\end{eqnarray}
where $\Pi_{h}(T)$ and $\Pi_{s}(T)$ are as in Eq.~\ref{sxyTherMh} and Eq.~\ref{sxyTherMs} respectively.
The tree level gauge boson masses are as in the case (1) above, except that the B-field has now a mass term for non-zero $s$ values. Obviously, in this case, the photon has a tree-level $s$-induced background mass term. The Z-boson-photon mass mixing term, which holds for both longitudinal and transverse components, at tree level is now 
\begin{eqnarray}
\label{sxyTreeMc2}
M^2_{Z,\gamma}(h,s) =  \left( \begin{array}{cc}  g^2&   - g g' \\
				                                       - g g'&   g'^2  \end{array} \right) \frac{h^2}{4} +
				                                     \left( \begin{array}{cc}   0&   0\\
				                                         0&   g'^2  \end{array} \right) s^2.
\end{eqnarray}
And finally the longitudinal gauge boson thermal masses are
\begin{eqnarray}
\label{sxyGMc2}
  \mathrm{M}^2_L(h,s,T) =    \mathrm{M}^2_L(h,T) + \left( \begin{array}{cccc}0 & 0      & 0    &   0  \\
				                                       0 & 0 & 0    &   0 \\
				                                       0 &   0   & 0&   0\\
				                                       0 &   0   & 0&   g'^2  \end{array} \right) s^2  , 
 \end{eqnarray}
where $ \mathrm{M}^2_L(h,T)$ is as for case (1) of Eq.~(\ref{sxyGM}).

\section{Relic density and annihilation cross sections in PNC Model} 
\label{AppAnn}

{
Here we give the relevant equations needed to evaluate the $\chi$ relic density for the PNC model, case(2), studied at the end of Sec.~\ref{PNCc2}.   

To obtain an estimate for the $\chi$ relic density we solve the usual DM Boltzmann (Lee-Weinberg) equation using the classic approximate method. 
Here, the freeze-out temperature $T_f$ is solved iteratively via the equation
\begin{eqnarray}
\label{xf}
x_f \equiv \frac{m_{\chi}}{T_f}  = \log{ \frac{m_{\chi}}{2 \pi^3} \sqrt{\frac{45 m^2_{\rm pl}}{2 g_{\star} x_f}} \langle \sigma v_{\rm rel} \rangle } ,
 \end{eqnarray}
where $m_{\rm pl}$ is the Planck mass and $g_{\star}(T)$ is the number of effective energy degrees of freedom for which we use the standard model values. 
For the thermally averaged annihilation cross section we use the expression \cite{Gondolo:1990dk}
\begin{eqnarray}
\label{annv}
 \langle \sigma v_{\rm rel} \rangle = \frac{1}{16 m^4_{\chi} T K^2_2(m_{\chi} / T)} \int^{\infty}_{4m^2_{\chi}}  s \sqrt{s-4m^2_{\chi}} \,K_1(\sqrt{s}/ T) \sigma v_{rel} \, \rm{d}s,
 \end{eqnarray}
where $K_1$ and $K_2$ are the modified Bessel functions of the second kind and the annihilation cross section to different final states $i$, $\sigma = \sigma_{\bar{\chi} \chi \rightarrow i \bar{i}}$, are given below in Eqs.~\ref{annf},\ref{anng},\ref{annh} (and below \ref{Logf}) and \ref{annhs}.
The relic density for the $\chi$ particle is obtained from equation
\begin{eqnarray}
\label{DMden}
\Omega h^2 = \frac{1.07 \times 10^9 x_f} {\sqrt{g_{\star}} m_{\rm pl}  \langle \sigma v_{\rm rel} \rangle }.
 \end{eqnarray}
The above approximate solution for the relic density does not work well near resonances, but for our purposes it yields a reasonable overall estimate of the DM density. We use it for the full range assumed by the $\chi$ masses that can yield a first order EWPT. 
A more accurate solution for the relic density can be obtain by solving the Lee-Weinberg equation numerically, however we do not expect this to affect our conclusions, that is, $\chi$ can be at most a subdominant DM component.  

Now we summarise the $\chi$ annihilation cross sections into different final states including SM fermions, Higgs and gauge bosons and the $\Phi$ scalar.

The annihilation cross section into fermionic final states  reads
\begin{eqnarray}
\label{annf}
\sigma_{\bar{\chi} \chi \rightarrow f \bar{f}} = \frac{ y_{\chi }^2 \sin ^2\omega  \cos ^2\omega \,  m_f^2 X_f }{4 \pi  v_{\text{EW}}^2}
 \left|  \frac{1}{s-m_{\Phi }^2+i \Gamma _{\Phi } m_{\Phi }}+\frac{1}{s-m_{\phi }^2+i \Gamma _{\phi }
   m_{\phi }} \right|^2 s \beta _{\chi } \beta _f^3 ,
\end{eqnarray}
where the factor $X_f = 1$ for leptons and for quarks
\begin{eqnarray}
 X_f = N_c  \left(1+ \frac{4 \alpha_s }{3 \pi } \left(\frac{3}{2} \log
   \left(\frac{m_f^2}{s}\right)+\frac{9}{4}\right) \right),
 \end{eqnarray}   
which takes into account the relevant one loop QCD correction for light quarks in the range $s \lesssim 90^2$ \cite{Drees:1990dq}. The color factor $N_c =3$ and we use the value $\alpha_s = 0.12$ for the strong coupling.  The velocity factor is 
$\beta _i = \sqrt{1-4 m_i^2 /s}$.

The annihilation cross section into gauge bosons is
\begin{eqnarray}
\label{anng}
\sigma_{\bar{\chi} \chi \rightarrow B \bar{B}}&=&f_B \frac{y_{\chi }^2 \sin ^2 \omega  \cos ^2 \omega m_B^4}{2 \pi 
   v_{\text{EW}}^2}  \left| \frac{1}{s-m_{\Phi }^2+i \Gamma _{\Phi } m_{\Phi
   }}+\frac{1}{s-m_{\phi }^2+i \Gamma _{\phi } m_{\phi }}\right|^2 \times \\
&&\beta _{\chi} \beta _B \left(\frac{s^2}{4
    m_B^4}-\frac{s}{m_B^2}+3\right),
 \end{eqnarray}
 where $B$ refers to $W$ or $Z$ bosons and the factor $f_W =1$ and $f_Z = 1/2$ respectively.

The trilinear scalar couplings, needed when calculating the tree level annihilation cross sections into scalar final states, are defined as
\begin{eqnarray}
\label{tricoup}
\lambda_{\phi \phi \phi } & \equiv & v \lambda_h \cos ^3\omega +u \lambda _{hs} \sin \omega  \cos ^2\omega +v \lambda_{hs} \sin ^2\omega  \cos \omega +u \lambda_s \sin ^3\omega  \\
\lambda _{\Phi \Phi \Phi } & \equiv  &-v \lambda _h \sin ^3\omega +u \lambda _{hs} \sin ^2\omega  \cos \omega -v \lambda_{hs} \sin \omega  \cos ^2\omega +u \lambda _s \cos ^3\omega  \\
\lambda _{\Phi \phi \phi }& \equiv  & 3 u \lambda _s \sin ^2\omega  \cos \omega -3 v \lambda _h \sin \omega  \cos ^2\omega  + \nonumber \\
& & \lambda _{hs} \left( u \cos ^3\omega -2 u \sin ^2\omega  \cos \omega - v \sin ^3\omega +2 v \sin \omega \cos ^2\omega \right) \\
\lambda _{\phi \Phi \Phi }   & \equiv  & 3 u \lambda _s \sin \omega  \cos ^2 \omega + 3 v \lambda _h \sin ^2\omega  \cos \omega + \nonumber \\
& & \lambda _{hs}  \left(v \cos ^3\omega -2 u \sin \omega  \cos^2 \omega + u \sin ^3\omega - v \sin^2 \omega  \cos \omega \right) ,
 \end{eqnarray}
where $v \equiv v_{\rm EW}$ and $u \equiv v_{\rm EW} \tan \omega$.

The annihilation cross section into the scalar $\phi$ (Higgs) final states is
\begin{eqnarray}
\label{annh}
\sigma_{\bar{\chi} \chi \rightarrow \phi \phi} &=& \frac{ y_{\chi}^2}{64 \pi  s^2 \beta_{\chi }^2} \Biggl[
2 y_{\chi }^2  \sin ^4 \omega  \left(\frac{2 \, \rm{Logf} \left(m_{\phi }^4+4 m_{\chi }^2 \left(s-4 m_{\chi
   }^2\right)\right)}{s-2 m_{\phi }^2}-s \beta _{\phi } \beta _{\chi }\right) + \nonumber \\
 && y_{\chi}^2 \sin ^4 \omega \, \rm{Logf} \left(-2 m_{\phi }^2+8
   m_{\chi }^2+s\right)+  \nonumber \\
 && 2 y_{\chi}^2 \sin ^4 \omega  \Biggl(  \frac{s \beta _{\phi } \beta _{\chi } \left(8 m_{\phi }^4-4 \left(8 m_{\chi
   }^2+s\right) m_{\phi }^2+64 m_{\chi }^4+s^2-s^2 \beta _{\phi }^2 \beta _{\chi }^2\right)}{s^2 \beta
   _{\phi }^2 \beta _{\chi }^2-\left(s-2 m_{\phi }^2\right)^2}+  \nonumber \\
&&  \rm{Logf} \left(-2 m_{\phi }^2+8 m_{\chi }^2+s\right) \Biggr)+  \nonumber \\
    &&  16 m_{\chi } y_{\chi}  \sin^2\omega  \left( \rm{Logf}  \left(2 m_{\phi }^2-8 m_{\chi }^2
 +s\right)+2 s \beta _{\phi } \beta _{\chi }\right) \times  \nonumber  \\
&& \Bigg( \frac{3 \sin \omega  \left(m_{\phi}^2-s\right) \lambda_{\phi \phi \phi }} {\left(m_{\phi }^2-s\right)^2+m_{\phi }^2 \Gamma _{\phi }^2} +
\frac{\cos \omega  \left(m_{\Phi }^2-s\right) \lambda _{\Phi \phi \phi }}{\left(m_{\Phi }^2-s\right)^2+m_{\Phi }^2 \Gamma _{\Phi }^2}  \Bigg) - \nonumber \\
&& 16 s \left(4 m_{\chi }^2-s\right) \beta _{\phi } \beta _{\chi } \Bigg(
    \frac{9 \sin ^2\omega  \lambda _{\phi \phi \phi }^2 }
   {\left(m_{\phi }^2-s\right)^2+m_{\phi }^2 \Gamma _{\phi }^2} 
    +\frac{\cos^2\omega  \lambda _{\Phi\phi \phi }^2}{ \left(m_{\Phi }^2-s\right)^2+m_{\Phi }^2 \Gamma _{\Phi }^2}  + \nonumber \\
 && \frac{6 \cos \omega 
   \sin \omega \lambda _{\Phi \phi \phi } \lambda _{\phi \phi \phi } \left(\left(m_{\phi }^2-s\right) \left(m_{\Phi }^2-s\right)+m_{\phi } m_{\Phi }
   \Gamma _{\phi } \Gamma _{\Phi }\right) }
   {\left(\left(m_{\phi }^2-s\right)^2+m_{\phi }^2 \Gamma _{\phi }^2\right)
   \left(\left(m_{\Phi }^2-s\right)^2+m_{\Phi }^2 \Gamma _{\Phi }^2\right)}   \Bigg) \Biggr],
 \end{eqnarray}
where 
 \begin{eqnarray}
 \label{Logf}
\rm{Logf} \equiv \log \left| \frac{s (\beta _{\phi } \beta _{\chi }+1)-2 m_{\phi }^2}{s \left(\beta _{\phi } \beta _{\chi
   }-1\right)+2 m_{\phi }^2} \right| .
 \end{eqnarray}

The annihilation cross section into heavier $\Phi$ scalars, $\sigma_{\bar{\chi} \chi \rightarrow \Phi \Phi}$, is obtained by interchanging $\sin \omega \leftrightarrow \cos \omega $ and $\phi  \leftrightarrow \Phi$ in the above cross section, $\sigma_{\bar{\chi} \chi \rightarrow \phi \phi}$, given in Eq.~\ref{annh}. 

Finally, the annihilation cross section into $\phi + \Phi$ final states reads
\begin{eqnarray}
\label{annhs}
\sigma_{\bar{\chi} \chi \rightarrow \phi \Phi} &=& \frac{1}{64 \pi  s^2 \beta _{\chi }^2} \Biggl[ 
\frac{ -4 y_{\chi }^4 \sin ^2(\omega ) \cos ^2(\omega )}{\left(m_{\Phi}^2+m_{\phi }^2-s\right) \left(-\beta _{\chi } \sqrt{\lambda _f}+m_{\Phi }^2+m_{\phi }^2-s\right)
   \left(\beta _{\chi } \sqrt{\lambda _f}+m_{\Phi }^2+m_{\phi }^2-s\right)} \times  \nonumber \\
 &&  \Biggl( \text{LogF} \left(-s \left(m_{\Phi
   }^2-16 m_{\chi }^2+3 m_{\phi }^2\right)+2 \left(m_{\phi }^2-4 m_{\chi }^2\right) \left(2 m_{\Phi
   }^2+4 m_{\chi }^2+m_{\phi }^2\right)+s^2\right) \times  \nonumber \\
&& \left(\beta _{\chi }^2 \lambda _f-\left(m_{\Phi}^2+m_{\phi }^2-s\right)^2\right) 
   - 2 \beta _{\chi } \sqrt{\lambda _f} \left(m_{\Phi}^2+m_{\phi }^2-s\right) \times  \nonumber \\ 
&& \Big( \left(\beta _{\chi }^2-3\right) m_{\phi }^4+\beta _{\chi }^2 \left(m_{\Phi
   }^2-s\right)^2+m_{\phi }^2 \left(-2 \beta _{\chi }^2 \left(m_{\Phi }^2+s\right)-2 m_{\Phi }^2+16
   m_{\chi }^2+3 s\right)+ \nonumber  \\
&&  s m_{\Phi }^2-m_{\Phi }^4-32 m_{\chi }^4-s^2 \Big) \Biggr)-  \nonumber \\
  &&16 m_{\chi } y_{\chi }^3
   \sin (\omega ) \cos (\omega ) \left(2 \beta _{\chi } \sqrt{\lambda _f}-\text{LogF} \left(-8 m_{\chi
   }^2+2 m_{\phi }^2+s\right)\right) \times \nonumber \\ 
  && \left(\frac{\lambda _{\Phi \phi \phi } \sin (\omega )
   \left(s-m_{\phi }^2\right)}{\Gamma _{\phi }^2 m_{\phi }^2+\left(m_{\phi
   }^2-s\right)^2}-\frac{\lambda _{\phi \Phi \Phi } \cos (\omega ) \left(m_{\Phi
   }^2-s\right)}{\Gamma _{\Phi }^2 m_{\Phi }^2+\left(m_{\Phi }^2-s\right)^2}\right) +\nonumber  \\
  && 8 \beta _{\chi }
   \sqrt{\lambda _f} y_{\chi }^2 \left(s-4 m_{\chi }^2\right) \Bigg(\frac{\lambda _{\phi \Phi \Phi }^2
   \cos ^2(\omega )}{\Gamma _{\Phi }^2 m_{\Phi }^2+\left(m_{\Phi }^2-s\right)^2}+ \frac{\lambda _{\Phi
   \phi \phi }^2 \sin ^2(\omega )}{\Gamma _{\phi }^2 m_{\phi }^2+\left(m_{\phi}^2-s\right)^2} + \nonumber \\
  && \frac{2 \lambda_{\Phi \phi \phi } \lambda _{\phi \Phi \Phi } \sin (\omega ) \cos (\omega ) \left(\Gamma _{\Phi }
   \Gamma _{\phi } m_{\Phi } m_{\phi }+\left(m_{\Phi }^2-s\right) \left(m_{\phi
   }^2-s\right)\right)}{\left(\Gamma _{\Phi }^2 m_{\Phi }^2+\left(m_{\Phi }^2-s\right)^2\right)
   \left(\Gamma _{\phi }^2 m_{\phi }^2+\left(m_{\phi }^2-s\right)^2\right)} \Bigg) \Biggr] ,
 \end{eqnarray}
where 
 \begin{eqnarray}
\rm{LogF} &\equiv& \log \left| \frac{m_{\Phi }^2+m_{\phi }^2-s +\beta _{\chi } \sqrt{\lambda _f} }{m_{\Phi }^2+m_{\phi }^2-s-\beta _{\chi } \sqrt{\lambda _f}} \right|, \\
\lambda _f &\equiv& \left(m_{\Phi }^2-s\right){}^2-2 m_{\phi }^2 \left(m_{\Phi }^2+s\right)+m_{\phi }^4.
 \end{eqnarray}

}


\clearpage 

\begin{thebibliography}{99}




\bibitem{Sakharov:1967dj}
  A.~D.~Sakharov,
  Pisma Zh.\ Eksp.\ Teor.\ Fiz.\  {\bf 5} (1967) 32
   [JETP Lett.\  {\bf 5} (1967) 24]
   [Sov.\ Phys.\ Usp.\  {\bf 34} (1991) 392]
   [Usp.\ Fiz.\ Nauk {\bf 161} (1991) 61].

\bibitem{Kajantie:1996mn}
  K.~Kajantie, M.~Laine, K.~Rummukainen and M.~E.~Shaposhnikov,
  Phys.\ Rev.\ Lett.\  {\bf 77} (1996) 2887
  [hep-ph/9605288].

\bibitem{Espinosa:2011ax}
  J.~R.~Espinosa, T.~Konstandin and F.~Riva,
  Nucl.\ Phys.\ B {\bf 854} (2012) 592
  [arXiv:1107.5441 [hep-ph]].
  


\bibitem{Cline:2009sn}
  J.~M.~Cline, G.~Laporte, H.~Yamashita and S.~Kraml,
  JHEP {\bf 0907} (2009) 040
  [arXiv:0905.2559 [hep-ph]].
  
\bibitem{Espinosa:2008kw}
  J.~R.~Espinosa, T.~Konstandin, J.~M.~No and M.~Quiros,
  Phys.\ Rev.\ D {\bf 78} (2008) 123528
  [arXiv:0809.3215 [hep-ph]].

\bibitem{Elias-Miro:2014pca}
  J.~Elias-Miro, J.~R.~Espinosa and T.~Konstandin,
  JHEP {\bf 1408} (2014) 034
  [arXiv:1406.2652 [hep-ph]].
 
  
\bibitem{Martin:2014bca}
  S.~P.~Martin,
  Phys.\ Rev.\ D {\bf 90} (2014) 016013
  [arXiv:1406.2355 [hep-ph]].

\bibitem{Veltman:1980mj}
  M.~J.~G.~Veltman,
  Acta Phys.\ Polon.\ B {\bf 12} (1981) 437.

\bibitem{Coleman:1973jx}
  S.~R.~Coleman and E.~J.~Weinberg,
  Phys.\ Rev.\ D {\bf 7} (1973) 1888.




\bibitem{Arnold:1992rz}
  P.~B.~Arnold and O.~Espinosa,
  Phys.\ Rev.\ D {\bf 47} (1993) 3546
   [Erratum-ibid.\ D {\bf 50} (1994) 6662]
  [hep-ph/9212235].

\bibitem{Fodor:1994bs}
  Z.~Fodor and A.~Hebecker,
  Nucl.\ Phys.\ B {\bf 432} (1994) 127
  [hep-ph/9403219].
 
\bibitem{Farakos:1994kx}
  K.~Farakos, K.~Kajantie, K.~Rummukainen and M.~E.~Shaposhnikov,
  Nucl.\ Phys.\ B {\bf 425} (1994) 67
  [hep-ph/9404201].
 


 
 
  
  
\bibitem{Rummukainen:1998as}
  K.~Rummukainen, M.~Tsypin, K.~Kajantie, M.~Laine and M.~E.~Shaposhnikov,
  Nucl.\ Phys.\ B {\bf 532} (1998) 283
  [hep-lat/9805013].

\bibitem{Laine:1998jb}
  M.~Laine and K.~Rummukainen,
  Nucl.\ Phys.\ Proc.\ Suppl.\  {\bf 73} (1999) 180
  [hep-lat/9809045].

\bibitem{Csikor:1998eu}
  F.~Csikor, Z.~Fodor and J.~Heitger,
  Phys.\ Rev.\ Lett.\  {\bf 82} (1999) 21
  [hep-ph/9809291].

\bibitem{Kajantie:1995kf}
  K.~Kajantie, M.~Laine, K.~Rummukainen and M.~E.~Shaposhnikov,
  Nucl.\ Phys.\ B {\bf 466} (1996) 189
  [hep-lat/9510020].


\bibitem{Kajantie:1996qd}
  K.~Kajantie, M.~Laine, K.~Rummukainen and M.~E.~Shaposhnikov,
  Nucl.\ Phys.\ B {\bf 493} (1997) 413
  [hep-lat/9612006].

\bibitem{Philipsen:1996af}
  O.~Philipsen, M.~Teper and H.~Wittig,
  Nucl.\ Phys.\ B {\bf 469} (1996) 445
  [hep-lat/9602006].

\bibitem{Gurtler:1997hr}
  M.~Gurtler, E.~M.~Ilgenfritz and A.~Schiller,
  Phys.\ Rev.\ D {\bf 56} (1997) 3888
  [hep-lat/9704013].

\bibitem{Karsch:1996yh}
  F.~Karsch, T.~Neuhaus, A.~Patkos and J.~Rank,
  Nucl.\ Phys.\ Proc.\ Suppl.\  {\bf 53} (1997) 623
  [hep-lat/9608087].


\bibitem{Linde:1980ts}
  A.~D.~Linde,
  Phys.\ Lett.\ B {\bf 96} (1980) 289.


\bibitem{Carrington:1991hz}
  M.~E.~Carrington,
  Phys.\ Rev.\ D {\bf 45} (1992) 2933.



\bibitem{Antipin:2013exa}
  O.~Antipin, M.~Mojaza and F.~Sannino,
  Phys.\ Rev.\ D {\bf 89} (2014) 085015
  [arXiv:1310.0957 [hep-ph]].




\bibitem{Coleman:1977py}
  S.~R.~Coleman,
  Phys.\ Rev.\ D {\bf 15} (1977) 2929
   [Phys.\ Rev.\ D {\bf 16} (1977) 1248].

\bibitem{Callan:1977pt}
  C.~G.~Callan, Jr. and S.~R.~Coleman,
  Phys.\ Rev.\ D {\bf 16} (1977) 1762.


\bibitem{Linde:1980tt}
  A.~D.~Linde,
  Phys.\ Lett.\ B {\bf 100} (1981) 37.
  
\bibitem{Linde:1981zj}
  A.~D.~Linde,
  Nucl.\ Phys.\ B {\bf 216} (1983) 421.
  


\bibitem{Choi:1993cv}
  J.~Choi and R.~R.~Volkas,
  Phys.\ Lett.\ B {\bf 317} (1993) 385
  [hep-ph/9308234].

\bibitem{Ham:2004cf}
  S.~W.~Ham, Y.~S.~Jeong and S.~K.~Oh,
  J.\ Phys.\ G {\bf 31} (2005) 857
  [hep-ph/0411352].

\bibitem{Ahriche:2007jp}
  A.~Ahriche,
  Phys.\ Rev.\ D {\bf 75} (2007) 083522
  [hep-ph/0701192].

\bibitem{Profumo:2007wc}
  S.~Profumo, M.~J.~Ramsey-Musolf and G.~Shaughnessy,
  JHEP {\bf 0708} (2007) 010
  [arXiv:0705.2425 [hep-ph]].

\bibitem{Ashoorioon:2009nf}
  A.~Ashoorioon and T.~Konstandin,
  JHEP {\bf 0907} (2009) 086
  [arXiv:0904.0353 [hep-ph]].

\bibitem{Ahriche:2012ei}
  A.~Ahriche and S.~Nasri,
  Phys.\ Rev.\ D {\bf 85} (2012) 093007
  [arXiv:1201.4614 [hep-ph]].


\bibitem{Cline:2013gha}
  J.~M.~Cline, K.~Kainulainen, P.~Scott and C.~Weniger,
  Phys.\ Rev.\ D {\bf 88} (2013) 055025
  [arXiv:1306.4710 [hep-ph]].
  
\bibitem{Cline:2012hg}
  J.~M.~Cline and K.~Kainulainen,
  JCAP {\bf 1301} (2013) 012
  [arXiv:1210.4196 [hep-ph]].
  
 

\bibitem{Espinosa:2011eu}
  J.~R.~Espinosa, B.~Gripaios, T.~Konstandin and F.~Riva,
  JCAP {\bf 1201} (2012) 012
  [arXiv:1110.2876 [hep-ph]].

\bibitem{Zeldovich:1974uw}
  Y.~B.~Zeldovich, I.~Y.~Kobzarev and L.~B.~Okun,
  Zh.\ Eksp.\ Teor.\ Fiz.\  {\bf 67} (1974) 3
   [Sov.\ Phys.\ JETP {\bf 40} (1974) 1].



\bibitem{Espinosa:2007qk}
  J.~R.~Espinosa and M.~Quiros,
  Phys.\ Rev.\ D {\bf 76} (2007) 076004
  [hep-ph/0701145].


\bibitem{Tamarit:2014dua}
  C.~Tamarit,
  Phys.\ Rev.\ D {\bf 90} (2014) 055024
  [arXiv:1404.7673 [hep-ph]].


\bibitem{Alanne:2014bra}
  T.~Alanne, K.~Tuominen and V.~Vaskonen,
  arXiv:1407.0688 [hep-ph].




\bibitem{Fairbairn:2013uta}
  M.~Fairbairn and R.~Hogan,
  JHEP {\bf 1309} (2013) 022
  [arXiv:1305.3452 [hep-ph]].

\bibitem{Li:2014wia}
  T.~Li and Y.~-F.~Zhou,
  arXiv:1402.3087 [hep-ph].

\bibitem{Frandsen:2013bfa}
  M.~T.~Frandsen, F.~Sannino, I.~M.~Shoemaker and O.~Svendsen,
  Phys.\ Rev.\ D {\bf 89} (2014) 055004
  [arXiv:1312.3326 [hep-ph]].


\bibitem{Ade:2015xua}
  P.~A.~R.~Ade {\it et al.}  [Planck Collaboration],
  arXiv:1502.01589 [astro-ph.CO].

\bibitem{Kainulainen:2015raa}
  K.~Kainulainen, K.~Tuominen and J.~Virkaj\"arvi,
  arXiv:1504.07197 [hep-ph].

\bibitem{Akerib:2013tjd}
  D.~S.~Akerib {\it et al.}  [LUX Collaboration],
  Phys.\ Rev.\ Lett.\  {\bf 112} (2014) 091303
  [arXiv:1310.8214 [astro-ph.CO]].

\bibitem{Giardino:2013bma}
  P.~P.~Giardino, K.~Kannike, I.~Masina, M.~Raidal and A.~Strumia,
  JHEP {\bf 1405} (2014) 046
  [arXiv:1303.3570 [hep-ph]].



\bibitem{Anderson:1991zb}
  G.~W.~Anderson and L.~J.~Hall,
  Phys.\ Rev.\ D {\bf 45} (1992) 2685.

\bibitem{Espinosa:1993bs}
  J.~R.~Espinosa and M.~Quiros,
  Phys.\ Lett.\ B {\bf 305} (1993) 98
  [hep-ph/9301285].

\bibitem{Barger:2008jx}
  V.~Barger, P.~Langacker, M.~McCaskey, M.~Ramsey-Musolf and G.~Shaughnessy,
  Phys.\ Rev.\ D {\bf 79} (2009) 015018
  [arXiv:0811.0393 [hep-ph]].
  
\bibitem{Gonderinger:2012rd}
  M.~Gonderinger, H.~Lim and M.~J.~Ramsey-Musolf,
  Phys.\ Rev.\ D {\bf 86} (2012) 043511
  [arXiv:1202.1316 [hep-ph]].
  
\bibitem{Jiang:2015cwa}
  M.~Jiang, L.~Bian, W.~Huang and J.~Shu,
  arXiv:1502.07574 [hep-ph].



\bibitem{Chikashige:1980ui}
  Y.~Chikashige, R.~N.~Mohapatra and R.~D.~Peccei,
  Phys.\ Lett.\ B {\bf 98} (1981) 265.

\bibitem{Kondo:1991jz}
  Y.~Kondo, I.~Umemura and K.~Yamamoto,
  Phys.\ Lett.\ B {\bf 263} (1991) 93.

\bibitem{Enqvist:1992va}
  K.~Enqvist, K.~Kainulainen and I.~Vilja,
  Nucl.\ Phys.\ B {\bf 403} (1993) 749.

\bibitem{Sei:1992np}
  N.~Sei, I.~Umemura and K.~Yamamoto,
  Phys.\ Lett.\ B {\bf 299} (1993) 286.
  



\bibitem{Farzinnia:2013pga}
  A.~Farzinnia, H.~J.~He and J.~Ren,
  Phys.\ Lett.\ B {\bf 727} (2013) 141
  [arXiv:1308.0295 [hep-ph]].
  
\bibitem{Farzinnia:2014xia}
  A.~Farzinnia and J.~Ren,
  Phys.\ Rev.\ D {\bf 90} (2014) 015019
  [arXiv:1405.0498 [hep-ph]].

\bibitem{Farzinnia:2014yqa}
  A.~Farzinnia and J.~Ren,
  arXiv:1408.3533 [hep-ph].


\bibitem{Dissauer:2012xa}
  K.~Dissauer, M.~T.~Frandsen, T.~Hapola and F.~Sannino,
  Phys.\ Rev.\ D {\bf 87} (2013) 3,  035005
  [arXiv:1211.5144 [hep-ph]].

\bibitem{DelNobile:2012tx}
  E.~Del Nobile, C.~Kouvaris, P.~Panci, F.~Sannino and J.~Virkajarvi,
  JCAP {\bf 1208} (2012) 010
  [arXiv:1203.6652 [hep-ph]].

\bibitem{Antipin:2013bya}
  O.~Antipin, J.~Krog, M.~Mojaza and F.~Sannino,
  Nuclear Physics, Section B (2014), pp. 125-134
  [arXiv:1311.1092 [hep-ph]].
  
\bibitem{Katz:2014bha}
  A.~Katz and M.~Perelstein,
  JHEP {\bf 1407} (2014) 108
  [arXiv:1401.1827 [hep-ph]].
  
\bibitem{Steigman:2012nb}
  G.~Steigman, B.~Dasgupta and J.~F.~Beacom,
  Phys.\ Rev.\ D {\bf 86} (2012) 023506
  [arXiv:1204.3622 [hep-ph]].
  
\bibitem{Gildener:1976ih}
  E.~Gildener and S.~Weinberg,
  Phys.\ Rev.\ D {\bf 13} (1976) 3333.

 
\bibitem{Dolan:1973qd}
  L.~Dolan and R.~Jackiw,
  Phys.\ Rev.\ D {\bf 9} (1974) 3320.
 
 
\bibitem{Drees:1990dq}
  M.~Drees and K.~i.~Hikasa,
  Phys.\ Lett.\ B {\bf 240} (1990) 455
   [Phys.\ Lett.\ B {\bf 262} (1991) 497].

\bibitem{Gondolo:1990dk}
  P.~Gondolo and G.~Gelmini,
  Nucl.\ Phys.\ B {\bf 360} (1991) 145.


\end{thebibliography}
\end{document}